\documentclass[aps,prd,reprint,superscriptaddress,nofootinbib]{revtex4-1}
\usepackage[all]{xy}
\usepackage{amsmath,amsthm,amssymb}
\usepackage[dvips]{graphicx}
\usepackage{comment}
\usepackage{array}
\usepackage{bm}
\allowdisplaybreaks[1]

\usepackage{xcolor}
\usepackage{hyperref}
\usepackage{tcolorbox}
\usepackage{mathrsfs}
\usepackage{float}

\newcommand{\Einf}{E_{\infty}}
\newcommand{\El}{{\sf El}}
\newcommand{\Linf}{L_{\infty}}
\newcommand{\chiinf}{\chi_{\infty}}

\begin{document}

\title{Self-force correction to the deflection angle in black-hole scattering: a scalar charge toy model}

\def\Soton{Mathematical Sciences, University of Southampton, 
Southampton SO17 1BJ, United Kingdom}

\author{Leor Barack}
\affiliation{\Soton}

\author{Oliver Long}
\affiliation{\Soton}

\date{\today}
\begin{abstract}
Using self-force methods, we consider the hyperbolic-type scattering of a pointlike particle carrying a scalar charge $Q$ off a Schwarzschild black hole. For given initial velocity and impact parameter, back-reaction from the scalar field modifies the scattering angle by an amount $\propto\! Q^2$, which we calculate numerically for a large sample of orbits (neglecting the gravitational self-force). Our results probe both strong-field and field-weak scenarios, and in the latter case we find a good agreement with post-Minkowskian expressions. The scalar-field self-force has a component tangent to the four-velocity that exchanges particle's mass with scalar-field energy, and we also compute this mass exchange as a function along the orbit. The expressions we derive for the scattering angle (in terms of certain integrals of the self-force along the orbit) can be used to obtain the gravitational self-force correction to the angle in the physical problem of a binary black hole with a large mass ratio. We discuss the remaining steps necessary to achieve this goal.
\end{abstract}

\maketitle

\section{Introduction} 

The deflection angle in hyperbolic black-hole scattering is a useful diagnostic of the two-body dynamics in general relativity. For example, information gleaned from post-Minkowskian (PM) calculations of the scattering angle provides a powerful calibration of the effective one-body (EOB) model of interacting black holes \cite{Damour:2016gwp,Damour:2017zjx,Damour2020,Bini:2020rzn}, in turn informing precision models of gravitational-wave sources for detector experiments. Further motivation to study black-hole scattering is provided by the recent direct link observed between scattering and bounds-orbit observables \cite{NeillRothstein2013,Damour:2017zjx,CheungRothstein2018,Kalin:2019rwq,KalinPorto2020,ChoKalin2022}, using effective-field-theory methods \cite{GoldbergRothstein2006}. The problem attracts considerable attention even outside the gravitational-physics community, with rapid progress being made through adaptation of well-developed methods from other areas of theoretical physics. A prime example are the emerging dictionaries that translate between quantum scattering amplitudes and classical gravitational dynamics (using advanced amplitude methods such as generalized unitarity \cite{BernDixon1994,BernDixon1995} and double copy \cite{KawaiLewellen1986,BernCarrasco2008,BernCarrasco2010}), leading in the past few years to a much accelerated development of the PM theory of gravitationally interacting binaries \cite{Bern:2019nnu,Bern:2019crd,Bern:2020buy,Bern:2020uwk}. Similar calculations have also been performed using effective-field-theory methods \cite{KalinPorto2020_PMEFT,Kalin:2020fhe,Liu:2021zxr,DlapaKalin2022,DlapaKalin2022_Cons,KalinNeef2022}. Thus the physical problem of black-hole scattering is today a lively arena for exchange and synergy between traditionally distinct fields of physics. Fundamentally, what makes this exchange possible is the relatively ``clean'' nature of the scattering scenario, where (in common with the analogous particle-physics problem) one has well-defined asymptotic `in' and `out' states of zero binding energy. 

So far, much of the work on black-hole scattering has been formulated in the context of PM theory, which is based on a weak-field approximation; at leading order the scattering trajectory is a straight line is flat space, and one seeks to incorporate the effect of gravitational interaction order by order in the gravitational constant $G$. Our goal here is to advance a complementary perturbative approach based on black-hole perturbation theory, which completely does away with the weak-field approximation, instead incorporating an expansion in the mass ratio $q(\leq 1)$ of the binary system. In this approach, the leading-order trajectory is a timelike geodesic in the exact spacetime of the larger object (say, a Kerr black hole), and one seeks to incorporate self-force and other post-geodesic terms order by order in $q$. One then has access to the full richness of the strong-field scattering dynamics, albeit at the cost of a priori restricting the validity of the analysis to small mass ratios. The complementary of the PM and self-force treatments has the benefit of allowing us to perform mutual validity checks, and also opens the possibility for the two approaches to inform each other in interesting ways. For example, it was noted by Damour in \cite{Damour2020} that the complete conservative 2-body dynamics through 4PM order can be inferred in full (i.e., for an arbitrary mass ratio) simply from first-order self-force calculations of the scattering angle. Similarly, and remarkably, a second-order self-force calculation would provide access to the full conservative dynamics through as high an order as 6PM.

In Ref.\ \cite{LongBarack2021} we initiated a program to calculate the scattering angle in the self-force approximation, without a PM expansion. In that work we developed and implemented a method for reconstructing the linear metric perturbation from a mass particle on a scattering orbit around a Schwarzschild black hole, in a gauge suitable for self-force calculations. The reconstruction procedure starts from a certain scalar-like Hertz potential that is obtained (numerically) by solving the (spin $\pm 2$) Teukolsky equation in the time domain. We have illustrated and tested the workability of the method with a simple time-domain numerical scheme, demonstrating the calculation of the metric perturbation and its derivatives along the orbit. We have not, in that paper, taken the extra steps of computing the back-reaction force on the orbit and from it the $O(q)$ correction to the geodesic scattering angle. Our purpose here is to carry out these extra steps, completing the (numerical) calculation of the scattering angle for strong-field orbits through $O(q)$. 

In the current paper we derive practical, ready-to-use formulas for the scattering angle through $O(q)$, expressed as functionals of self-force components along the orbit. We provide expressions for the full self-force effect, as well as---to enable comparison with PM results---for the conservative and dissipative effects in separate. We present two equivalent formulations using two different parametrizations of the scattering orbit.  Our first formulation uses the eccentricity $e$ and semilatus rectum $p$ as orbital parameters, with the associated radial phase serving as integration variable along the orbit, while our second formulation utilizes the radial coordinate as a parameter along (each of the two, in/outbound legs of) the orbit. The two methods are of course equivalent, but we opt to present them both here, as each can have different advantages under different numerical implementation schemes. We also obtain the leading-order PM reduction of our equations, to enable comparison with existing PM results. 

We then carry on to present a full numerical implementation, for both our formulations. Here, however, we take a sideways step in our program and consider a simpler physical model, in which the small mass particle is replaced with a scalar charge. In this toy model the role of the linear metric perturbation is played by the scalar field sourced by the charge (which we take to satisfy the minimally coupled Klein-Gordon equation on the fixed Schwarzschild geometry of the large black hole), and the role of the gravitational self-force is played by the back-reaction force from the scalar field; in our model, the gravitational self-force itself is neglected. Our scattering-angle formulation applies unaltered, simply replacing the gravitational self-force with the component of the scalar-field self-force orthogonal to the charge's four-velocity. (The self-force component tangent to the four-velocity, which we will also calculate, has the effect of exchanging rest mass with scalar-field energy; see Section \ref{Sec:Scalar}.) We numerically solve the scalar-field equation with the appropriate sourcing term in the time domain, construct the self-force using standard mode-sum regularization, and then apply our integral formulas to compute the scattering angle, as corrected by the self-force, for a range of orbital parameters. Our numerical method works best for strong-field orbits, but we are able to probe sufficiently into the weak-field domain to enable us to test our results against the leading-order PM expressions available from Ref.\ \cite{GrallaLobo2022}. We find a reassuring agreement. 

The main purpose of our detour through a scalar-field toy model is to enable us to check our scattering-angle formulation in a cleaner environment, and without yet having to give due consideration to the additional subtleties inherent in the gravity case, primarily those surrounding gauge ambiguity. 
In addition, the simple numerical method we have applied in Ref.\ \cite{LongBarack2021} to compute the metric perturbation is highly suboptimal, and a change of methodology is necessary to enable accurate scattering-angle calculations in the gravity case. In our concluding section here we elaborate on the necessary steps to improve the numerical method and describe our current efforts in that direction. 

The structure of this paper is as follows. In Sec.\ \ref{Sec:intro} we review scattering orbits and the derivation of the scattering angle in the geodesic limit $q\to 0$. Sec.\ \ref{Sec:SFMotion} analyzes the particle's equations of motion under the effect of the leading-order gravitational self-force, in the physical pure-gravity problem.  In particular, we derive the self-force correction (defined with fixed initial velocity $v$ and impact parameter $b$) to the orbit's periastron distance, eccentricity $e$ and semilatus rectum $p$. In Sec.\ \ref{Sec:SFchi} we derive a formula for the self-force correction to the scattering angle (again defined with fixed $v$ and $b$) as a functional of self-force components, expressed in terms of an integral over the relativistic anomaly $\chi$ of the orbit; and in Sec.\ \ref{Sec:SFr} we derive an alternative formula using the $v,b$ parametrization directly, with the radius as integration variable. Section \ref{sec:PMlimit} describes the PM expansion of our formulas, with a comparison to existing analytical results. 

Section \ref{Sec:Scalar} then presents our scalar-charge toy model, reviews the calculation of the scalar-field self-force via mode-sum regularization, and discusses the PM reduction of our scattering-angle formulas in the scalar case. As a by-product, we analytically derive the leading (3PM) dissipative term of the scattering angle for the scaler-charge model. The following two sections present a full numerical implementation using a time-domain finite-difference code based on characteristic coordinates: Sec.\ \ref{Sec:ScalarImplementation} describes our numerical method (with much of the detail delegated to Appendix \ref{app:FDS}), and in Sec.\ \ref{Sec:results} we display and analyze a sample of our numerical results. Section \ref{Sec:conc} contains a summary and a discussion of the extension to gravity.

Throughout this work we use natural geometrized units, with $G=1=c$, and adopt the metric signature $({-}{+}{+}{+})$.  The large central object is taken to be a Schwarzschild black hole with mass $M$ and spacetime metric  $ds^2=-f^{-1}(r)dt^2 +f(r) dr^2 +r^2(d\theta^2+\sin^2\theta d\varphi^2)$,
where $f(r):=1-2M/r$. The small object is a pointlike particle of mass $\mu\ll M$ and (in our scalar-field model) carrying a scalar charge $Q$ such that $Q^2\ll M\mu$,  the reason for which requirement to be made clear in Section \ref{Sec:Scalar}. The scattering trajectory of the particle on the Schwarzschild background is described by $x^\alpha = x_p^\alpha(\tau)$, with tangent four-velocity $u^\alpha = dx_p^\alpha/d\tau$, where $\tau$ is proper time along the orbit (setting $\tau=0$ at the periastron point). Without loss of generality we take the trajectory to lie in the equatorial plane of a fixed Schwarzschild coordinate system, so that in that frame it is described by $x^\alpha=\big(t_p(\tau),r_p(\tau),\pi/2,\varphi_p(\tau)\big)$.

\section{Scattering angle in the geodesic limit}
\label{Sec:intro}

In the limit $q=\mu/M\to 0$, the scattering process reduces to geodesic motion on a Schwarzschild background.  The geodesic equations of motion can be written in a first-integral form, 
\begin{eqnarray}
\dot{t}_p&=& E/f(r_p), \label{tdot}\\
\dot{\varphi}_p &=& L/r_p^2 ,\label{phidot}\\
\dot{r}_p &=& \pm \sqrt{E^2-V(r_p;L)},  \label{rdot}
\end{eqnarray}
where (recall) $f(r)=1-2M/r$, an overdot denotes $d/d\tau$, and the radial effective potential is
\begin{equation}\label{V}
V(r;L)=f(r)\left(1+L^2/r^2\right).
\end{equation}
$E:=-u_{t}$ and $L:=u_{\varphi}$ are the test particle's energy and angular momentum per $\mu$, constants of the geodesic motion. For hyperbolic orbits we have 
\begin{equation}\label{gamma}
E=(1-v^2)^{-1/2} >1 ,
\end{equation} 
where $v:=[\dot{r}_p^2+(r\dot\varphi_p)^2]^{1/2}/\dot{t}_p\Big\vert_{\tau\to -\infty}$ is the magnitude of the initial 3-velocity (with respect to time $t$), and $E$ is then the initial ``gamma factor'' of the incident particle. 
The particle actually scatters back to infinity (and does not fall into the black hole) only if $L>L_{\rm crit}(E)$, where the critical value of the angular momentum is the relevant simultaneous solution of $\partial_r V(r;L)=0$ and $E^2=V(r;L)$:
\begin{equation}
L_{\rm crit}(E) = \frac{M}{vE}\sqrt{(27E^4+9\alpha E^3-36 E^2-8\alpha E+8)/2},
\end{equation}
where $\alpha:=\sqrt{9E^2-8}$.

The {\it impact parameter} of the scattering geodesic is defined as 
\begin{equation}
b:=\lim_{\tau\to-\infty} r_p\sin\left|\varphi_p(\tau)-\varphi_p(-\infty)\right|,
\end{equation}
which, using (\ref{phidot}) and (\ref{rdot}), gives
\begin{equation}\label{b}
b= \frac{L}{\sqrt{E^2-1}} = \frac{L}{v E}.
\end{equation}
For a scattering orbit we need $b>b_{\rm crit}$, where $b_{\rm crit}:=L_{\rm crit}(E)/(vE)$. It can be checked that $b_{\rm crit}(E)$ is a monotonically decreasing function. Thus the minimal possible value of the impact parameter is
\begin{equation}
\lim_{E\to\infty}b_{\rm crit}(E) = 3\sqrt{3}M \simeq 5.196 M .
\end{equation}
Incident particles with $b<3\sqrt{3}M$ are captured by the black hole for any $E$.

As orbital parameters for the family of scattering geodesics we can use either pairs $\{E,L\}$ or $\{v,b\}$, with the conversion obtained using Eqs.\ (\ref{gamma}) and (\ref{b}). We note $v$ and $b$ are attributes of the initial state of the scattering process (both are defined via the limit $\tau\to -\infty$). They will therefore remain useful parameters even as (in subsequent sections) we add in self-force effects and the motion no longer admits conserved energy and angular momentum.

\subsection{Perisatron distance}

For given $E>1$ and $L>L_{\rm crit}(E)$, the cubic equation $\dot{r}_p^2=E^2-V(r;L)=0$ admits three real roots $r=\{r_1,r_2,r_0\}$, satisfying $r_1<0$ and $2M<r_2<r_0$.  They are given explicitly by \cite{MaartenUnpublished} 
\begin{eqnarray}\label{r0}
r_0&=&\frac{6M}{1-2\zeta \sin\left(\frac{\pi}{6}-\xi\right)},\\ \label{r1}
r_1&=&\frac{6M}{1-2\zeta \sin\left(\frac{\pi}{6}+\xi\right)},\\ \label{r2}
r_2&=&\frac{6M}{1+2\zeta \cos \xi},
\end{eqnarray}
with
\begin{eqnarray}
\label{eqn:zetaxi}
\zeta&:=& \sqrt{1-12M^2/L^2},\nonumber\\
\xi&:=& \frac{1}{3}\arccos\left(\frac{1+(36-54E^2)M^2/L^2}{\zeta^3} \right).
\end{eqnarray}
The largest of these roots, $r_0$, is the periastron distance, i.e.\ the radius of nearest approach to the black hole. Even though only the turning point $r_0$ is physically relevant in our scattering problem, the values $r_1$ and $r_2$ will play a role in our self-force formulation in Sec.\ \ref{Sec:SFr}.
For later use, we note here the relation 
\begin{equation}\label{r3}
r_2= \frac{2M r_0 r_1}{r_0 r_1-2M(r_0 + r_1)}.
\end{equation}

The periastron distance decreases with increasing $E$ (at fixed $L$), down to the ``light ring,'' $r_0\to 3M$, for $E\to\infty$. At fixed $E$ (or $v$), $r_0$ increases with $L$ (or b). It is instructive to consider the asymptotic form of $r_0$ at large impact parameter: Substituting for $E,L$ in terms of $v,b$ in Eq.\ (\ref{r1}) and then expanding in powers of $1/b$ at fixed $v$, we find
\begin{equation}\label{rminPM}
r_0=b - \frac{M}{v^2}+\left(\frac{1-4v^2}{2v^4}\right) \frac{M^2}{b}
+O(b^{-2}).
\end{equation}
Thus $r_0 \sim b$ at large $b$, as long as $v$ is not too small. When $v$ is small, a large impact parameter $b$ does not necessarily imply ``weak field''; as an extreme example, consider the zero-binding-energy zoom-whirl orbit studied in Ref.\ \cite{Baracketal2019}, which has $v=0$ and $b\to\infty$, yet $r_0=4M$. The form of Eq.\ (\ref{rminPM}) motivates the choice 
\begin{equation}\label{eqn:v^2b}
\frac{M}{v^2 b} \ll 1
\end{equation}
as our definition of the ``weak-field'' domain of the scattering problem. We shall come back to this in Sec.\ \ref{sec:PMlimit}, when we compare our numerical self-force results to PM expressions. 


\subsection{The $e,p$ parametrization}

We note that any two of the roots $\{r_0,r_1,r_2\}$ can provide an alternative parametrization of the scattering geodesics, in lieu of $\{E,L\}$ or $\{v,b\}$. From the periastron distance $r_0$ and the (negative) root $r_1$ one can construct a convenient, geometrically motivated parametrization in terms of an eccentricity $e(>1)$ and an (a-dimentionalized) semilatus rectum $p$, defined through
\begin{equation}\label{rminpe}
r_0= \frac{M p}{1+e}, \quad\quad r_1= \frac{M p}{1-e} ,
\end{equation}
analogous to their bound-orbit definitions. From (\ref{r3}), the third root is then given by
\begin{equation}\label{r1r2r3_ep}
r_2 = \frac{2Mp}{p-4},
\end{equation}
which, we note, does not depend on $e$. The conversion relations between $\{e,p\}$ and $\{E,L\}$ can be obtained using Eqs.\ (\ref{r0})--(\ref{r2}), and work out to be the same as they are for bound orbits:
\begin{equation}
E^2 =
\frac{(p-2)^2-4e^2}{p(p-3-e^2)}, \quad\quad
L^2 =
\frac{p^2M^2}{p-3-e^2}.
\label{ELep}
\end{equation} 
To invert the relations (\ref{ELep}) entails solving cubic equations, and the results are cumbersome. But it is relatively simple to express $(e,p)$ in terms of $(L,r_0)$ [where $r_0$ itself can be obtained from $(E,L)$ using Eq.\ (\ref{r0})]:
\begin{equation}\label{eofL}
e=\frac{L^2 r_0-2Mr_0^2+\sqrt{L^4r_0f(r_0)(r_0+6M)-16M^2 L^2r_0^2}}{2M(L^2+r_0^2)},
\end{equation}
with $p=(r_0/M)(1+e)$ from Eq.\ (\ref{rminpe}).

The main advantage of the $e,p$ parametrization is that it allows us to describe the radial motion in the simple, Keplerian-like form
\begin{equation}\label{rofchi}
r_p(\chi)=\frac{Mp}{1+e\cos\chi}.
\end{equation}
The radial phase $\chi$ is a relativistic anomaly along the orbit, taking the values $\chi\in (-\chi_\infty,\chi_\infty)$ with 
\begin{equation}\label{chiinf}
\chi_\infty=\arccos(-1/e) ,
\end{equation}
and with periastron passage corresponding to $\chi=0$. 
The relation between $t_p$ and $\chi$ is found using Eqs.\ (\ref{tdot})--(\ref{rdot}) and then substituting from (\ref{rofchi}) and (\ref{ELep}):
\begin{align}
\frac{dt_p}{d\chi} =  
\frac{\dot{t}_p}{\dot{r}_p}\frac{dr_p}{d\chi}
= &\: \frac{Mp^2}{(p-2-2e\cos\chi)(1+e\cos\chi)^2}
\nonumber\\
& \times \sqrt{\frac{(p-2)^2-4e^2}{p-6-2e\cos\chi}}.
\label{eq:dt_dchi} 
\end{align}

\subsection{Scattering angle}

An expression  $\varphi_p(\chi)$ along the orbits can be found by integrating 
\begin{equation}
\frac{d\varphi_p}{d\chi} =\frac{\dot{\varphi}_p}{\dot{r}_p}\frac{dr}{d\chi}=
\sqrt{\frac{p}{p-6-2e\cos\chi}},
\label{eq:dphi_dchi}
\end{equation}
where we have used (\ref{tdot})--(\ref{rdot}) and then substituted from (\ref{ELep}) and (\ref{rofchi}).
This equation has an explicit integral in terms of an Elliptic function: 
\begin{equation}\label{phiofchi}
\varphi_p(\chi)=\varphi_p(0)+k\sqrt{p/e}\, \El_1\Big(\frac{\chi}{2};-k^2\Big),
\end{equation}
where 
\begin{equation}\label{k}
k=2\sqrt{\frac{e}{p-6-2e}},
\end{equation}
and $\El_1$ is the incomplete elliptic integral of the first kind:
\begin{equation}\label{eqn:El1}
\El_1(\varphi;k)=\int_0^{\varphi} (1-k\sin^2 x)^{-1/2}dx.
\end{equation}

From (\ref{phidot}) we see that $\dot\varphi_p\to 0$ for $r\to\infty$. Therefore $\varphi_p\to$ const for $\chi\to\pm\chi_\infty$. Let $\varphi_{\rm in}$ and $\varphi_{\rm out}$ represent the asymptotic values of $\varphi_p$ for $\chi\to -\chi_\infty$ and $\chi\to\chi_\infty$, respectively. From Eq.\ (\ref{phiofchi}), the difference between them is given by 
\begin{eqnarray}
\Delta\varphi &:=&
\varphi_{\rm out}-\varphi_{\rm in}  \nonumber\\
&=& k\sqrt{p/e}\left[\El_1 \Big(\frac{\chi_\infty}{2};-k^2\Big)-\El_1\Big(-\frac{\chi_\infty}{2};-k^2\Big)\right]\nonumber\\
&=& 2k\sqrt{p/e}\, \El_1\Big(\frac{\chi_\infty}{2};-k^2\Big).
\label{Deltaphi}
\end{eqnarray}
The scattering angle is defined as 
\begin{equation}\label{delta}
\psi:= \Delta\varphi-\pi. 
\end{equation}
It can be checked that $\psi\to 0$ in the PM limit $v^2b \to \infty$ [cf.\ Eq.\ (\ref{deltaphiPM}) below], and that $\psi\to\infty$ in the ``zoom-whirl'' limit $p\to 6+2e$ (equivalent to $b\to b_{\rm crit}$).




\subsection{PM expansion} \label{subsec:PM}

For our PM analysis in Sec.\ \ref{sec:PMlimit} we will need the weak-field reduction of some of the above geodesic-limit expressions.  In what follows we present the relevant PM expansions, working at the order required to obtain the first subleading PM term of the scattering angle $\psi$, which is the order at which self-force terms first occur. 

First, consider the weak-field form of the eccentricity $e$. In Eq.\ (\ref{eofL}) we replace $L\to b v(1-v^2)^{-1/2}$ [recalling (\ref{gamma}) and (\ref{b})], substitute the PM expansion of $r_0$ from Eq.\ (\ref{rminPM}), and then re-expand in $M/b$ at fixed $v$. The result is
\begin{equation}\label{ePM}
e=v^2\frac{b}{M}+\left(\frac{1-4v^2-8v^4}{2v^2}\right)\frac{M}{b}+O\left(\frac{M}{b}\right)^3.
\end{equation}
This also gives, recalling Eq.\ (\ref{chiinf}),
\begin{eqnarray}\label{chiinfPM}
\chi_\infty =
 \frac{\pi}{2}+\frac{1}{v^2}\, \frac{M}{b}
+O\left(\frac{M}{b}\right)^3.\nonumber\\
\end{eqnarray}
We can now use $p=r_0(1+e)/M$ with the expansions (\ref{rminPM}) and (\ref{ePM}) to obtain
\begin{equation}\label{pPM}
p=v^2 \frac{b^2}{M^2}-4(1+v^2) +O\left(\frac{M}{b}\right)^2.
\end{equation}
Note $e\propto b$ and $p\propto b^2$ at large $b$.

Substituting the expansions (\ref{ePM}) and (\ref{pPM}) in Eq.\ (\ref{k}) now gives
\begin{align}\label{kPM}
k= 2\sqrt{\frac{M}{b}}\Big[ 
1+\frac{M}{b} + \left(\frac{1+16v^2+6v^4}{4v^4}\right)\frac{M^2}{b^2} 
\nonumber \\
+O\left(\frac{M}{b}\right)^3 \Big].
\end{align}
The elliptic function in Eq.\ (\ref{Deltaphi}) can be expanded in its index $-k^2$ about $k=0$, giving 
\begin{eqnarray}
\El_1\Big(\frac{\chi_\infty}{2};-k^2\Big)&=&
\frac{1}{2}\chi_\infty-\frac{1}{8}\left(\chi_\infty-\sin\chi_\infty\right)k^2 \nonumber\\
&&+\frac{3}{256}[6\chi_\infty-8\sin\chi_\infty+\sin(2\chi_\infty)]k^4 
\nonumber\\
&& +O(k^6).
\end{eqnarray}
Putting everything together in Eq.\ (\ref{delta}) we finally obtain
\begin{align}\label{deltaphiPM}
\psi\ = \frac{2(1+v^2)}{v^2}\frac{M}{b}+\frac{3\pi(4+v^2)}{4v^2}\frac{M^2}{b^2}
+O\left(\frac{M}{b}\right)^3 .
\end{align}
This agrees with the geodesic limit of standard PM expressions [compare, for example, with Eq.\ (6) of \cite{GrallaLobo2022}].



\section{Motion with first-order self-force}
\label{Sec:SFMotion}

We proceed to consider the equations of motion with a leading-order gravitational self-force term. (The case of a scalar-field self-force is closely analogous; it will be discussed separately in Sec.\ \ref{Sec:Scalar}.)  We thus now endow the particle with mass $\mu\ll M$, define the mass ratio $q:=\mu/M\ll 1$, and henceforth use $q$ for order counting. The mass $\mu$ sources a perturbation of the Schwarzschild geometry associated with $M$, whose linear piece exerts a gravitational self-force $\mu q F^{\alpha}$, where $F^{\alpha}$ is the leading-order self-acceleration per $q$. Our ultimate goal is to calculate the resulting $O(q)$ correction to the scattering angle $\delta$ away from its geodesic value [for fixed $(v,b)$]. In this section, as a preparatory step, we will derive the corrections to the periastron distance $r_0$, eccentricity $e$ and semilatus rectum $p$ [all at fixed $(v,b)$].

The equation of self-forced motion reads
\begin{equation}\label{EoM_SF}
u^\beta \nabla_\beta u^\alpha = q F^\alpha ,
\end{equation}
where $u^\alpha$ is now the tangent four-velocity along the perturbed orbit, normalized such that $g_{\alpha\beta}u^\alpha u^\beta=-1$, where $g_{\alpha\beta}$ is the background Schwarzschild metric. $\nabla_\beta$ denotes the covariant derivative compatible with $g_{\alpha\beta}$, and tensor indices are raised and lowered using $g_{\alpha\beta}$ throughout our discussion. Again we introduce spherical coordinates and take the orbit to lie in its equatorial plane, which, from symmetry, we can do without loss of generality even with self-force. Equation (\ref{EoM_SF}) then takes the explicit form
\begin{eqnarray}
\dot{E}&=& - q F_t \label{tdotF}\\
\dot{L} &=& q F_\varphi \label{phidotF}\\
\ddot{r}_p &=& -\frac{1}{2}\frac{\partial V(r_p;L)}{\partial r_p}+q F^r,  \label{rdotdotF}
\end{eqnarray}
where $E(\tau):=-u_t$ and $L(\tau):=u_\varphi$ are no longer conserved but are now functions along the orbit. 
The normalization condition (\ref{rdot}) still applies, with the replacements $E\to E(\tau)$ and $L\to L(\tau)$:
\begin{equation}\label{rdot_SF}
\dot{r}_p(\tau) = \pm \sqrt{E(\tau)^2-V(r_p(\tau);L(\tau))}.
\end{equation} 

The self-force along the geodesic scattering orbit can be split into conservative and dissipative pieces,
\begin{equation}
F^\alpha = F^\alpha_{\rm cons}+F^\alpha_{\rm diss},
\end{equation}
unambiguously defined, respectively, from the ``half retarded plus half advanced'' and the ``half retarded minus half advanced'' linear metric perturbations. In practice, it is often simpler to construct the conservative and dissipative pieces using the special symmetry of Kerr geodesics. Specifically for our equatorial scattering geodesics, and (recall) taking $\tau=0$ at periastron, we have 
\begin{eqnarray}\label{F_cons_diss}
F^\alpha_{\rm cons}(\tau) &=& \frac{1}{2}\left[F^\alpha(\tau)\pm F^\alpha(-\tau)\right],
\nonumber \\
F^\alpha_{\rm diss}(\tau) &=& \frac{1}{2}\left[F^\alpha(\tau)\mp F^\alpha(-\tau)\right],
\end{eqnarray}
with the upper sign for $\alpha=r$ and the lower sign for $\alpha=t,\varphi$. Thus, in practice, the dissipative and conservative pieces can be constructed by appropriately combining the values of the self-force at two ``opposite'' points of the orbit, i.e.~ones with the same $r_p$ but opposite $\dot{r}_p$.

Given the full self-force $F^\alpha$, Eqs.\ (\ref{tdotF}) and (\ref{phidotF}) can be integrated immediately  to give
\begin{equation}
\label{eqn:ELwithSF}
E(\tau) = \Einf + q\Delta E(\tau) ,\quad\quad
L(\tau) = \Linf + q\Delta L(\tau) ,
\end{equation}
where 
\begin{eqnarray}
\Einf&:=& E(\tau\to-\infty) = (1-v^2)^{-1/2},\nonumber\\
\Linf&:=& L(\tau\to-\infty) = bv(1-v^2)^{-1/2},
\end{eqnarray}
and 
\begin{equation}\label{DeltaEL}
\Delta E(\tau):= - \int_{-\infty}^{\tau} F_t\, d\tau,\quad\quad
\Delta L(\tau):= \int_{-\infty}^{\tau} F_\varphi\, d\tau .
\end{equation}
The quantities $\Delta E(\tau)$ and $\Delta L(\tau)$ describe the self-force-induced change in the energy and angular momentum away from their initial values. The part of this change due to $F^\alpha_{\rm diss}$ accounts for radiative losses through gravitational radiation. There are generally also nonzero contributions to $\Delta E(\tau)$ and $\Delta L(\tau)$ due to $F^\alpha_{\rm cons}$  (which, however, integrate to zero for $\tau\to\infty$). 

\subsection{Self-force correction to $r_0$ }\label{subsec:r_p}

For fixed $(v,b)$, the self-force causes an $O(q)$ displacement in the periastron radius $r_0$, which we now derive. We let $\tilde r_0(v,b,q)$ represent the perturbed value of $r_0$, and write 
\begin{equation}
\tilde r_0= r_0 + q\, \delta r_0,
\end{equation}
where $r_0=r_0(v,b)$ is the geodesic value and $q\, \delta r_0(v,b)$ is the self-force correction. More precisely,
\begin{eqnarray}
r_0&: =& \lim_{q\to 0} \tilde r_p(v,b,q),\nonumber \\
\delta r_0&: =& \lim_{q\to 0}\frac{\partial \tilde r_0}{\partial q},
\end{eqnarray}
where the limits are taken with fixed $(v,b)$. 
The geodesic value $r_0(v,b)$ is given in Eqs.\ (\ref{r0}) [with Eq.\ (\ref{eqn:zetaxi}), replacing $E\to (1-v^2)^{-1/2}$ and $L\to bv(1-v^2)^{-1/2}$]. 

To obtain $\delta r_0$, we impose $\dot r_p(\tau(\tilde r_0))=0$ in Eq.\ (\ref {rdot_SF}), to obtain
\begin{equation}\label{normalization_sf}
E(\tau(\tilde r_0))^2 = V(\tilde r_0,L(\tau(\tilde r_0))).
\end{equation}
The linear perturbation of this equation with respect to $q$ is 
\begin{equation}\label{rminEq}
2\Einf \Delta E_0 = \frac{\partial V(r,L)}{\partial r}\bigg\vert_0 \delta r_0
+\frac{\partial V(r,L)}{\partial L}\bigg\vert_0 \Delta L_0 ,
\end{equation}
where the partial derivatives are evaluated at $(r,L)=(r_0,\Linf)$, and 
\begin{eqnarray} \label{Ep}
\Delta E_0 &:=& \Delta E(\tau(r_0))=- \int_{-\infty}^{0} F_t \, d\tau,
\\ \label{Lp}
\Delta L_0 &:=& \Delta L(\tau(r_0))= \int_{-\infty}^{0} F_\varphi\, d\tau.
\end{eqnarray}
Solving Eq.\ (\ref{rminEq}) for $\delta r_0$ gives
\begin{equation}\label{rmin1}
\delta r_0 = \frac{r_0(r_0-2M)\Linf \Delta L_0 - r_0^4\Einf\Delta E_0}{\Linf^2(r_0-3M)-Mr_0^2} .
\end{equation}

Equation (\ref{rmin1}) describes the shift in the coordinate location of the periastron [at fixed $(v,b)$] in terms of integrals of self-force components. This result will be needed in our derivation of the scattering angle in Sec.\ \ref{Sec:SFr}. 

\subsection{Self-force corrections to $p$ and $e$}
\label{Sec:epwithSF}

We can represent the self-force-perturbed radial motion again in the form (\ref{rofchi}), i.e., 
\begin{equation}\label{rofchi_sf}
\tilde r_p(\chi)=\frac{M\tilde p}{1+\tilde e\cos\chi},
\end{equation}
where overtildes denote perturbed values. Choosing the parameter $\chi$ such that $\chi=0$ at the periastron of the perturbed orbit, we have $\tilde r_0=M\tilde p/(1+\tilde e)$, and thus, from Eq.\ (\ref{normalization_sf}),
\begin{equation}\label{rdot_rmin_sf}
E_0^2 = V\left(\frac{M\tilde p}{1+\tilde e},L_0\right),
\end{equation}
where $E_0 := E(\tau(\tilde r_0))=\Einf+q\Delta E_0$ and similarly for $L_0$. This gives one relation between $(\tilde p,\tilde e)$ and $(E_0,L_0)$. As a second relation, to fully specify $(\tilde p,\tilde e)$ in terms of self-force integrals, we make the convenient choice
\begin{equation}\label{rdot_rmax_sf}
E_0^2 = V\left(\frac{M\tilde p}{1-\tilde e},L_0\right).
\end{equation}
Since, with this choice, the perturbed $(\tilde e,\tilde p)$ are related to $(E_0,L_0)$ exactly as $(e,p)$ were related to $(E,L)$ in the geodesic case, these relations are described explicitly by Eqs.\ (\ref{ELep}) with only the simple replacements $E\to E_0$ and $L\to L_0$:
\begin{equation}
E_0^2 =
\frac{(\tilde p-2)^2-4\tilde e^2}{\tilde p(\tilde p-3-\tilde e^2)}, \quad\quad
L_0^2 =
\frac{\tilde p^2M^2}{\tilde p-3-\tilde e^2}.
\label{ELinfep}
\end{equation} 
We emphasize that the definition of eccentricity and semilatus rectum for the perturbed orbit is a matter of choice. Our choice here is convenient in that it preserves the form of the relations (\ref{ELep}) (with the conserved geodesic $E,L$ replaced with their values at the periastron of the perturbed orbit).

We now write 
\begin{equation}
\tilde e=e+q\, \delta e, \quad\quad\, \tilde p=p +q\, \delta p ,
\end{equation}
where as usual the perturbation is defined for fixed $(v,b)$. The perturbations $\delta e$ and $\delta p$ are determined by varying Eqs.\ (\ref{ELinfep}) with respect to $q$ at fixed $(\Einf,\Linf)$ [and hence at fixed $(v,b)$]:
\begin{eqnarray}\label{Evar}
2\Einf \Delta E_0 &=& \frac{\partial}{\partial p}\left(\frac{(p-2)^2-4e^2}{p(p-3-e^2)}\right) \delta p\nonumber\\
&&+\frac{\partial}{\partial e}\left(\frac{(p-2)^2-4e^2}{p(p-3-e^2)}\right)\delta e, \quad\\
2\Linf \Delta L_0 &=& \frac{\partial}{\partial p}\left(\frac{p^2M^2}{p-3-e^2}\right)\delta p\nonumber\\
&&+\frac{\partial}{\partial e}\left(\frac{p^2M^2}{p-3-e^2}\right)\delta e. \label{Lvar}
\end{eqnarray}
Solving (\ref{Evar}) and (\ref{Lvar}) simultaneously for $\delta p$ and $\delta e$, we obtain
\begin{equation}\label{p1}
\delta p=\frac{2(p-3-e^2)}{(p-6)^2-4e^2}\left[
\frac{(p-4)^2}{pM^2}\Linf \Delta L_0 -p^2 \Einf \Delta E_0 \right],
\end{equation}
\begin{eqnarray}\label{e1}
\delta e&=&\frac{p-3-e^2}{e[(p-6)^2-4e^2]}
\Bigg[p(p-6-2e^2) \Einf \Delta E_0 
\nonumber\\
&& + \frac{(e^2-1)[(p-2)(p-6)+4e^2]}{p^2M^2}\Linf \Delta L_0 \Bigg].
\end{eqnarray}

Equations (\ref{p1}) and (\ref{e1}) describe the corrections to $e$ and $p$ [at fixed $(v,b)$] in terms of integrals of self-force components. These results will be needed in our derivation of the scattering angle in Sec.\ \ref{Sec:SFchi}.

\section{Self-force correction to the scattering angle $\psi$ }
\label{Sec:SFchi}

In what follows we parametrize the perturbed orbit using the pair $(\tilde p,\tilde e)$, with $\chi$ running along the orbit. We have $\chi=0$ at periastron, and $\chi\to \pm\tilde{\chi}_\infty=\pm \arccos(-1/\tilde e) $ for $t\to\pm\infty$. We think of $\tilde r_p(\tau)$ and $\tilde \varphi_p(\tau)$ now as functions of $\chi$ along the orbit. We have
\begin{align}
\frac{d\tilde\varphi_p}{d\chi}& =\frac{\dot{\tilde \varphi}_p}{\dot{\tilde r}_p}\frac{d\tilde r_p}{d\chi}
\nonumber\\
& = \frac{L(\chi)}{\tilde r_p(\chi)^2\sqrt{E(\chi)^2-V(\tilde r_p(\chi),L(\chi))}}\frac{M \tilde p \tilde e |\sin\chi|}{(1+\tilde e\cos\chi)^2},
\label{dphi_dchi_sf}
\end{align}
where $\tilde r_p(\chi)$ is given in Eq.\ (\ref{rofchi_sf}), and we henceforth think of $E$ and $L$ too as function of $\chi$ instead of $\tau$. It proves convenient to write
\begin{eqnarray}\label{ELchi}
E(\chi)&=&\Einf +q\Delta E(\chi) = E_0 + q(\Delta E(\chi)-\Delta E_0),
\nonumber\\
L(\chi)&=&\Linf +q \Delta L(\chi) = L_0 + q(\Delta L(\chi)-\Delta L_0),
\end{eqnarray}
recalling that subscripts `$0$' and `$\infty$' denote values at periastron and at $t\to -\infty$, respectively. 
Substituting (\ref{ELchi}) in (\ref{dphi_dchi_sf}), expanding in $q$, and then using Eq.\ (\ref{ELinfep}) to substitute for $E_0,L_0$ in terms of $\tilde e,\tilde p$, we find, of course, that the $O(q^0)$ term has the same form as in the geodesic case, Eq.\ (\ref{eq:dphi_dchi}). But there is now an $O(q)$ correction coming from the $O(q)$ terms in (\ref{ELchi}). Altogether we find
\begin{eqnarray}
\frac{d\tilde \varphi_p}{d\chi} &=&
\sqrt{\frac{\tilde p}{\tilde p-6-2\tilde e\cos\chi}} + q f_E(\chi;p,e)\left(\Delta E(\chi)-\Delta E_0\right)\nonumber\\
&&+ q f_L(\chi;p,e)\left(\Delta L(\chi)-\Delta L_0\right),
\label{dphi_dchi_sf2}
\end{eqnarray}
where 
\begin{eqnarray}\label{fEfL}
f_E &=& -\frac{p\sqrt{p-3-e^2}\sqrt{(p-2)^2-4e^2}}{e^2\sin^2\chi\, (p-6-2e\cos\chi)^{3/2}}, \nonumber\\
f_L &=& \frac{\sqrt{p-3-e^2}}{M\sqrt{p}\, e^2\sin^2\chi\, (p-6-2e\cos\chi)^{3/2}} \nonumber\\
&& \times\left[e^2(p-6)+p-2+2e(p-3-e^2)\cos\chi\right].\ \
\end{eqnarray}
In the $O(q)$ terms of (\ref{dphi_dchi_sf2}) we have replaced $(\tilde p,\tilde e)\to (p,e)$, the difference being of only $O(q^2)$.

The total accumulated orbital phase of the perturbed orbit is 
\begin{eqnarray}\label{Deltavarphi_sf}
\widetilde{\Delta\varphi} &=& \int_{-\tilde\chi_\infty}^{\tilde\chi_\infty}\frac{d\tilde \varphi_p}{d\chi}d\chi 
 = 2\tilde k\sqrt{\tilde p/\tilde e}\, \El_1\Big(\frac{\tilde\chi_\infty}{2};-\tilde k^2\Big)\nonumber\\
&&-q \int_{-\chiinf}^{\chiinf} d\chi 
\Bigg[
f_E(\chi)\int_{0}^{\chi} F_t(\chi')\tau_{\chi'}d\chi' \nonumber\\
&& \qquad \qquad\qquad- f_L(\chi)\int_{0}^{\chi} F_\varphi(\chi') \tau_{\chi'}d\chi' \Bigg]  ,
\end{eqnarray}
where we have recalled Eqs.\ (\ref{Deltaphi}) and (\ref{DeltaEL}), and have again dropped terms of $O(q^2)$. The Jacobian $\tau_{\chi}:=d\tau/d\chi$ can be evaluated along the background geodesic:
\begin{equation}\label{tauchi}
\tau_\chi = \frac{Mp \sqrt{p(p-3-e^2)}}{(1+e\cos\chi)^2\sqrt{p-6-2e\cos\chi}}.
\end{equation}
To obtain this we have used Eq.\ (\ref{eq:dt_dchi}) together with (\ref{tdot}), (\ref{rofchi}) and (\ref{ELep}).

We write the perturbed scattering angle $\tilde \psi:=\widetilde{\Delta\varphi}-\pi$ as
\begin{equation}\label{phisplit}
\tilde\psi = \psi + q\, \delta\psi,
\end{equation}
where the split between background and perturbation is, as always, defined with fixed $(v,b)$. The background function $\psi(e,p)$ was given in Eq.\ (\ref{delta}) [with (\ref{Deltaphi})].
The self-force correction $\delta\psi$ is obtained by taking the linear perturbation of (\ref{Deltavarphi_sf}) with respect to $q$ at fixed $(v,b)$:
\begin{eqnarray}\label{deltavarphi_sf}
\delta\psi&=&
\frac{\partial}{\partial p}\left[2k\sqrt{p/e}\, \El_1\Big(\frac{\chi_\infty}{2};-k^2\Big)\right]\delta p\nonumber\\
&&+\frac{\partial}{\partial e}\left[2k\sqrt{p/e}\, \El_1\Big(\frac{\chi_\infty}{2};-k^2\Big)\right]\delta e\nonumber \\
&&- \int_{-\chiinf}^{\chiinf} d\chi 
\Bigg[
f_E(\chi)\int_{0}^{\chi} F_t(\chi')\tau_{\chi'}d\chi' \nonumber\\
&& \qquad\qquad\qquad- f_L(\chi)\int_{0}^{\chi} F_\varphi(\chi') \tau_{\chi'}d\chi' \Bigg],
\end{eqnarray}
where $\chiinf(e)$ and $k(e,p)$ are given in Eqs.\ (\ref{chiinf}) and (\ref{k}), and where the perturbations $\delta p$ and $\delta e$ are those obtained above in Sec.\ \ref{Sec:epwithSF}, expressed in terms of the self-force integrals $\Delta E_0$ and $\Delta L_0$ [cf.\ Eqs.\ (\ref{p1}) and (\ref{e1})]. 

It is prudent to ask, at this point, whether the double integral in Eq.\ (\ref{deltavarphi_sf}) is actually convergent. The manifest $\sim(\sin\chi)^{-2}$ singularity of the functions $f_E$ and $f_L$ at the periastron ($\chi=0$) should raise a concern. To avoid distraction, we delegate answering this question to Appendix \ref{App:Convergence}. We show there that (i) the integrals of the individual $f_E$ and $f_L$ terms indeed fail to converge at $\chi=0$ (they each diverge there logarithmically in $\chi$, in general), but (ii) the full integral in Eq.\ (\ref{deltavarphi_sf}) is in fact convergent and well defined. The cancellation of the singularity between the two terms owes itself, essentially, to the normalisation relation $u^\alpha F_{\alpha}=0$. See Appendix \ref{App:Convergence} for details. 

\subsection{Simplification of Eq.\ (\ref{deltavarphi_sf}) for $\delta\psi$}

We now bring Eq.\ (\ref{deltavarphi_sf}) to a simpler, more ready-to-use form involving only a single integral over self-force components. 
First, we note that, in the first two lines of Eq.\ (\ref{deltavarphi_sf}), the coefficients of $\delta p$ and $\delta e$ can be written explicitly in terms of elliptic integrals, using the identities (valid for arbitrary $\phi,\kappa$)
\begin{align}
\frac{\partial\El_1(\phi;\kappa )}{\partial\phi}&=\frac{1}{\sqrt{1-\kappa \sin^2\phi}},
\nonumber\\
\frac{\partial\El_1(\phi;k)}{\partial \kappa}=
&\frac{1}{2\kappa(\kappa-1)}\Bigg[\frac{\kappa \cos\phi\, \sin\phi}{\sqrt{1-\kappa\sin^2\phi}}
\nonumber\\
& \qquad -(\kappa-1)\El_1(\phi,\kappa)-\El_2(\phi,\kappa) 
\Bigg] ,
\end{align}
where 
\begin{equation}\label{eqn:El2}
\El_2(\phi;\kappa)=\int_0^{\phi} (1-\kappa\sin^2 x)^{1/2}dx
\end{equation}
is the (incomplete) elliptic integral of the second kind. Substituting for $\delta p$ and $\delta e$ from Eqs.\ (\ref{p1}) and (\ref{e1}), the sum of the first two lines of (\ref{deltavarphi_sf}) then takes the form
\begin{equation}
\alpha_E(e,p) \Einf \Delta E_0 + \alpha_L(e,p) \Linf \Delta L_0,
\end{equation}
where, we find,
\begin{widetext}
\begin{align}\label{alphaE}
\alpha_E = \frac{2(p-3-e^2)p^{3/2}}{e^2(p-6+2e)^2(p-6-2e)^{3/2}}
\Bigg[
-(p-6)(p-6+2e)\El_1\Big(\frac{\chi_\infty}{2};-k^2\Big)
+(p^2-12p+12e^2+36) \El_2\Big(\frac{\chi_\infty}{2};-k^2\Big)
\nonumber\\
+\frac{16e^4-(p-6)^2(p-4)+4e^2(p^2-11p+24)}{\sqrt{(e^2-1)(p-4)(p-2e-6)}}
\Bigg],
\end{align}
\begin{align}\label{alphaL}
\alpha_L = \frac{2(p-3-e^2)}{ M^2 e^2 p^{3/2}(p-6+2e)^2(p-6-2e)^{3/2}}
\Bigg[
(p-6+2e)\left[(p-2)(p-6)+e^2(p^2-8p+24)-4e^4\right]\El_1\Big(\frac{\chi_\infty}{2};-k^2\Big)
\nonumber\\
+\left[-(p-2)(p-6)^2-e^2(p-2)(p^2-24)+4e^4(p-6)
\right] \El_2\Big(\frac{\chi_\infty}{2};-k^2\Big)
\nonumber\\
+\sqrt{\frac{(e^2-1)(p-4)}{p-6-2e}}\left[-(p-2)(p-6)^2-2e^2(p-4)(p+6)+8e^4\right]
\Bigg].
\end{align}
\end{widetext}

The third and fourth lines of (\ref{deltavarphi_sf}) involve double integrals of the self-force, which would make numerical evaluation inconvenient.  We can do away with this using integration by parts. To this end, we define 
\begin{equation}\label{calF}
{\cal F}_E(\chi):= 
\int_{\pm\chiinf}^\chi f_E(\chi') d\chi' ,
\quad
{\cal F}_L(\chi):= \int_{\pm \chiinf}^\chi f_L(\chi') d\chi',
\end{equation}
with $+$ sign for $\chi>0$ and with $-$ sign for $\chi<0$. These functions are well defined for all $\chi\ne 0$, and diverge (as $\sim \chi^{-1}$) in the limits $\chi\to 0^\pm$. Since $f_E$ and $f_L$ are bounded for $\chi\to\pm\chiinf$, we have
\begin{equation}\label{inflimit}
\lim_{\chi\to\pm\chiinf} {\cal F}_E(\chi) =0,
\quad\quad
\lim_{\chi\to\pm\chiinf} {\cal F}_L(\chi) =0;
\end{equation}
and since $f_E$ and $f_L$ are symmetric under $\chi\to -\chi$, we also have that ${\cal F}_E(\chi)$ and ${\cal F}_E(\chi)$ are antisymmetric:
\begin{equation}\label{antisym}
{\cal F}_E(-\chi) = -{\cal F}_E(\chi), \quad\quad {\cal F}_L(-\chi) = -{\cal F}_L(\chi).
\end{equation}
Integrating by parts in Eq.\ (\ref{deltavarphi_sf}), we write the sum of the third and fourth lines as 
\begin{align}
-&\left({\cal F}_E(\chi) \int_0^\chi  F_t(\chi')\tau_{\chi'} d\chi'
-{\cal F}_L(\chi) \int_0^\chi  F_\varphi(\chi')\tau_{\chi'} d\chi'\right)\Bigg\vert_{-\chiinf}^{\chiinf} \nonumber\\
&+\int_{-\chiinf}^{\chiinf} \Big({\cal F}_E(\chi') F_t(\chi')
-{\cal F}_L(\chi') F_\varphi(\chi')\Big)\tau_{\chi'} d\chi' ,
\end{align} 
and observe that the boundary terms all vanish by virtue of (\ref{inflimit}).

Collecting the above results, Eq.\ (\ref{deltavarphi_sf}) becomes  
\begin{align}\label{deltaphi1_almost_final}
\delta\psi=&
\alpha_E(e,p) \Einf \Delta E_0
+ \alpha_L(e,p) \Linf \Delta L_0 \nonumber\\
&+\int_{-\chiinf}^{\chiinf} \left[{\cal F}_E(\chi) F_t(\chi)
-{\cal F}_L(\chi) F_\varphi(\chi)\right]\tau_{\chi} d\chi,
\end{align}
or, recalling Eqs.\ (\ref{Ep}) and (\ref{Lp}),
\begin{align}\label{deltaphi1_final}
\delta\psi=
\int_{-\chiinf}^{\chiinf} \left[{\cal G}_E(\chi) F_t(\chi)
-{\cal G}_L(\chi) F_\varphi(\chi)\right]\tau_{\chi} d\chi,
\end{align}
where
\begin{eqnarray}
{\cal G}_E(\chi) &=& {\cal F}_E(\chi) -\alpha_E \Einf \Theta(-\chi), \\
{\cal G}_L(\chi) &=& {\cal F}_L(\chi) -\alpha_L \Linf \Theta(-\chi).
\end{eqnarray}
Here $\Theta(\cdot)$ is the standard Heaviside step function.
The functions ${\cal F}_E(\chi)$ and ${\cal F}_L(\chi)$, defined in Eq.\ (\ref{calF}), can be written explicitly in terms of incomplete elliptic functions of the first and second kind (but the expressions are cumbersome and we will not give them here). The constants $\alpha_E$ and $\alpha_L$ are given in Eqs.\ (\ref{alphaE}) and (\ref{alphaL}) explicitly in terms of incomplete elliptic integrals. 

It should be noted that the separate integrals over the $F_t$ and $F_\varphi$ terms in (\ref{deltaphi1_final}) do not individually converge, due to the $\sim \chi^{-1}$ singularity of ${\cal F}_E(\chi)$ and ${\cal F}_L(\chi)$ at the periastron; it is only the sum of two terms for which the integral converges. This follows from a similar analysis to the one we carry out in Appendix \ref{App:Convergence}. 

Equation (\ref{deltaphi1_final}) is our final expression for the full self-force correction $\delta\psi$. We will implement it numerically (for a scalar-field model) in Secs.\ \ref{Sec:ScalarImplementation} and \ref{Sec:results} of this work. 

\subsection{Conservative and dissipative pieces}
\label{subsec:SFchiCons&Diss}

It is often useful to consider the conservative and dissipative effects of the self-force in isolation. We can split $\delta\psi$ into a conservative contribution $\delta\psi_{\rm cons}$ and a dissipative contribution $\delta\psi_{\rm diss}$, defined by replacing $F_\alpha$ in (\ref{deltaphi1_final}) with $F_{\alpha}^{\rm cons}$ or $F_{\alpha}^{\rm diss}$, respectively. Recalling 
Eq.\ (\ref{F_cons_diss}), and our choice $\chi=0$ at the periastron, we note the symmetries
\begin{eqnarray}
F_\alpha^{\rm cons}(\chi) &=& - F_\alpha^{\rm cons}(-\chi),\\
F_\alpha^{\rm diss}(\chi) &=& F_\alpha^{\rm diss}(-\chi),
\end{eqnarray}
for $\alpha=t,\varphi$. Using this, and recalling also Eq.\ (\ref{antisym}), we observe that the product ${\cal F}_E(\chi)F_t^{\rm cons}(\chi)$ in Eq.\ (\ref{deltaphi1_final}) is symmetric under $\chi\to-\chi$, while ${\cal F}_E(\chi)F_t^{\rm diss}(\chi)$ is antisymmetric (and similarly for the $F_\varphi$ term). Therefore, in $\delta\psi_{\rm cons}$ we can fold the integral $\int_{-\infty}^0$ over onto $\int_{0}^\infty$, and in $\delta\psi_{\rm diss}$ the contribution from the ${\cal F}_E(\chi)$ and ${\cal F}_L(\chi)$ terms completely cancels out. We find
\begin{equation}\label{deltaphi_methodI_cons}
\delta\psi_{\rm cons}=
\int_{0}^{\chiinf} \left({\cal G}_E^{\rm cons} F_t^{\rm cons}
-{\cal G}_L^{\rm cons} F_\varphi^{\rm cons}\right)\tau_{\chi} d\chi,
\end{equation}
where
\begin{eqnarray}\label{calG_cons}
{\cal G}_E^{\rm cons}(\chi) &=& 2{\cal F}_E(\chi) +\alpha_E \Einf, \nonumber\\
{\cal G}_L^{\rm cons}(\chi) &=& 2{\cal F}_L(\chi) +\alpha_L \Linf,
\end{eqnarray}
and 
\begin{equation}\label{deltavarphi_diss}
\delta\psi_{\rm diss}=
\int_{0}^{\chiinf} \left(-\alpha_E\Einf F_t^{\rm diss}
+\alpha_L \Linf F_\varphi^{\rm diss}\right)\tau_{\chi} d\chi.
\end{equation}

Equations (\ref{deltaphi_methodI_cons}) and (\ref{deltavarphi_diss}) usefully prescribe the construction of $\delta\psi_{\rm cons}$ and $\delta\psi_{\rm diss}$ as integrals over the outgoing leg of the scattering orbit, given $F_{\alpha}^{\rm cons}$ and $F_{\alpha}^{\rm diss}$. Note however that, in practice, in our method, we would need to calculate the self-force along both legs of the orbit in order to construct $F_{\alpha}^{\rm cons}$ and $F_{\alpha}^{\rm diss}$ themselves. 

It may be observed, interestingly, that $\delta\psi_{\rm diss}$ can be written in a simple way in terms of the total energy and angular momentum (per $\mu q$) radiated in gravitational waves,
\begin{eqnarray}
E_{\rm rad}&=&-\int_{-\chiinf}^{\chiinf} F_t^{\rm diss}\tau_{\chi} d\chi
= -2\int_{0}^{\chiinf} F_t^{\rm diss}\tau_{\chi} d\chi, \nonumber\\
L_{\rm rad}&=&\int_{-\chiinf}^{\chiinf} F_\varphi^{\rm diss}\tau_{\chi} d\chi 
=2\int_{0}^{\chiinf} F_\varphi^{\rm diss}\tau_{\chi} d\chi\ ;
\end{eqnarray} 
specifically, we have
\begin{equation}
\label{deltavarphi_diss_fluxes}
\delta\psi_{\rm diss} = \frac{1}{2}\left(\alpha_E \Einf E_{\rm rad} + \alpha_L \Linf L_{\rm rad}\right).
\end{equation}
An analogous result has been obtained in post-Newtonian theory \cite{BiniDamour2012}, and also in PM theory \cite{BiniDamour2021}.

\section{Self-force correction to $\delta\psi$: an alternative method}
\label{Sec:SFr}

We present here an alternative way of calculating $\delta\psi$, where we avoid $(e,p)$ and parametrize directly in terms of $(v,b)$. Orbital integration is done with respect to $r$ instead of $\chi$. The method is somewhat more direct, as it skips the cumbersome step of converting between the $(e,p)$ and $(v,b)$ parametrizations. We present both formulations here since they can each be useful in different circumstances or when using different implementation methods, and since comparison can provide useful checks on the calculation.
 
\subsection{Scattering angle as a radial integral}

We start by rewriting Eq.\ (\ref{rdot}), for the geodesic case, in the form 
\begin{equation}
\dot{r}_p^2= (E^2-1)(r_p-r_0)(r_p-r_1)(r_p-r_2)/r_p^3.
\end{equation}  
The three roots on the right-hand side were given explicitly in Eqs.\ (\ref{r0})--(\ref{r2}).
The (geodesic) scattering angle is then
\begin{equation}\label{methodII_deltaphi0}
\psi = 2 \int_{r_0}^\infty (\dot\varphi_p/\dot r_p) dr -\pi = 2 \int_{r_0}^\infty \frac{H(r)}{\sqrt{r-r_0}} dr -\pi,
\end{equation}
where
\begin{equation}\label{H(r)}
H(r)=H(r;E,L) = \frac{L}{\sqrt{(E^2-1)r(r-r_1)(r-r_2)}}.
\end{equation}
Note $H(r)$ is smooth on the entire integration domain in Eq.\ (\ref{methodII_deltaphi0}). The integrand falls off as $\sim r^{-2}$ at $r\to\infty$, and it diverges like $(r-r_0)^{-1/2}$ at $r\to r_0$.  

Now consider the self-force-perturbed orbit. $E:=-u_t$ and $L:=u_\varphi$ now become slow functions along the orbit, which we gave explicitly (in terms of the self-force) in Eqs.\ (\ref{eqn:ELwithSF}).   Consequently, $r_0(E,L)$, $r_1(E,L)$ and $r_2(E,L)$ also become slow functions along the orbit; here $r_0(E,L)$ (and similarly for $r_1,r_2$) represents the {\it same functional relation} as in the geodesic case [Eqs.\ (\ref{r0})--(\ref{r2})], but with the arguments $E,L$ now being the self-force-corrected, slowly varying quantities.   

In order to use $r$ as a parameter along the orbit, we must consider the inbound and outbound legs separately. We use the notation $E^-(r)$ and $E^+(r)$ to denote the slowly-varying $E$ along the inbound and outbound legs, respectively, and we similarly introduce $L^{\pm}(r)$, $r_0^\pm(E(r),L(r))$, etc. Note $r_0^\pm(r)$ (the periastra of the tangent geodesics) are functions along the orbit, distinct from the constant self-force-perturbed value $\tilde r_0:= \min_{r}(r_0^{\pm}(r))=r_0+q\, \delta r_0$ calculated in Sec.\ \ref{subsec:r_p}. We have the relations $\tilde r_0= r_0^+(E^+(r_0),L^+(r_0))= r_0^-(E^-(r_0),L^-(r_0))$. We shall assume that $r$ is a monotonically decreasing function of $t$ (or $\tau$) on $-\infty< t < t_p(r_0)$, and a monotonically increasing function of $t$ (or $\tau$)  on $t_p(r_0)< t < \infty$; since this is true in the geodesic limit, it must also be true for sufficiently small $q$, from continuity.

With these notations, the total self-force-perturbed scattering angle is
\begin{equation}\label{methodII_deltaphi}
\tilde\psi = \sum_{\pm}\int_{\tilde r_0}^\infty \frac{\dot{\tilde \varphi}_p^\pm}{|\dot{\tilde r}_p^\pm|}\, dr -\pi = \sum_{\pm}\int_{\tilde r_0}^\infty \frac{\tilde H^\pm(r)}{\sqrt{r-r_0^\pm(r)}} dr -\pi,
\end{equation}
where
\begin{equation}
\tilde H^\pm(r)= \frac{L^\pm(r)}{\sqrt{\left[E^\pm(r)^2-1\right]r(r-r_1^\pm(r))(r-r_2^\pm(r))}}.
\end{equation}
Here $r_{1,2}^\pm(r)$ represent $r_{1,2}^\pm(E(r),L(r))$.


To obtain the perturbation $\delta\psi$ we need to vary the integral in (\ref{methodII_deltaphi}) with respect to $q$ at fixed $(v,b)$, or,  equivalently, at fixed $(\Einf,\Linf)$.  For this we would need to evaluate the derivatives of the integral with respect to $E(r)$, $L(r)$, $r_0$ and $r_{1,2}$. Varying with respect to $r_0$ is subtle, because of the singularity at the perisatron. To overcome this complication, we first integrate by parts:
\begin{align}\label{ibp}
\tilde\psi = & 
\sum_{\pm} \bigg[2 \sqrt{r-r_0^\pm(r)}\, \tilde H^\pm(r)\Big|_{\tilde r_0}^\infty
\nonumber\\
& -2 \int_{\tilde r_0}^\infty \sqrt{r-r_0^\pm(r)}\, \frac{d\tilde H^\pm(r)}{dr}dr 
\nonumber\\
& +\int_{\tilde r_0}^\infty \frac{\tilde H^\pm(r)}{\sqrt{r-r_0^\pm(r)}}\frac{dr_0^\pm}{dr}dr \bigg]
-\pi.
\end{align}
The functions $\tilde H^\pm(r)$ are bounded at $r=\tilde r_0$, and fall off like $r^{-3/2}$ at infinity, so the surface terms in Eq.\ (\ref{ibp}) vanish. We are left with
\begin{align}\label{deltavarphi_full}
\tilde\psi =& 
\sum_{\pm}\bigg[ -2 \int_{\tilde r_0}^\infty \sqrt{r-r_0^\pm(r)}
\nonumber\\
& \times \bigg(
\frac{\partial \tilde H^\pm}{\partial r} 
 \mp \frac{\partial H}{\partial \Einf} \frac{q F^\pm_t(r)}{|\dot{r}_p|} 
 \pm \frac{\partial H}{\partial \Linf} \frac{q F^\pm_\varphi(r)}{|\dot{r}_p|}
\bigg)dr 
\nonumber\\
&  +q\int_{r_0}^\infty \frac{H_0(r)}{\sqrt{r-r_0}} 
\left(\mp\frac{\partial r_0}{\partial \Einf}F^\pm_t \pm \frac{\partial r_0}{\partial \Linf}F^\pm_\varphi
\right)\frac{dr}{|\dot{r}_p|} \bigg] 
\nonumber\\
& -\pi ,
\end{align}
where $\partial_r$ is taken with fixed $(E,L)$, $\partial_E$ is taken with fixed $(r,L)$, and $\partial _L$ is taken with fixed $(r,E)$.
We have used $dE^\pm/dr = q \, d\Delta E^\pm/dr = \mp q F^\pm_t/|\dot{r}_p|$ and $dL^\pm/dr =  q \, d\Delta L^\pm/dr = \pm q F^\pm_\varphi/|\dot{r}_p|$, and replaced $r_0^\pm\to r_0(\Einf,\Linf)$ and $\tilde H^\pm\to H(r,\Einf,\Linf)$ (the geodesic relations) where such replacements amount only to omitting $O(q^2)$ terms in $\tilde\psi$.  The function $r_0=r_0(\Einf,\Linf)$ is the geodesic relation given in Eq.\ (\ref{r0}), with the replacements $E\to\Einf$ and $L\to \Linf$.
The geodesic limit of the expression in Eq.\ (\ref{deltavarphi_full}) is
\begin{equation}
\psi = -4 \int_{r_0}^\infty \sqrt{r-r_0}\, \frac{\partial H}{\partial r}\, dr -\pi,
\end{equation}
which, it can be checked, is equivalent to the expression in Eq.\ (\ref{methodII_deltaphi0}).

\subsection{Self-force correction $\delta\psi$} 

Varying $\tilde\psi$ in Eq.\ (\ref{deltavarphi_full}) with respect to $q$ at fixed $\Einf,\Linf$, we obtain
\begin{widetext}
\begin{align}\label{delta_varphi_v2}
\delta\psi =&
\sum_{\pm}\bigg[
\int_{r_0}^\infty \frac{1}{\sqrt{r-r_0}}\frac{\partial H}{\partial r}
\left(\frac{\partial r_0}{\partial \Einf}\Delta E^\pm(r)+ \frac{\partial r_0}{\partial \Linf}\Delta L^\pm(r)
\right)\, dr 
\nonumber\\
&  -2 \int_{r_0}^\infty \sqrt{r-r_0}
\left( 
  \frac{\partial^2 H}{\partial r\partial\Einf} \Delta E^\pm(r) 
  + \frac{\partial^2 H}{\partial r\partial\Linf} \Delta L^\pm(r) 
   \mp \frac{\partial H}{\partial\Einf} \frac{F^\pm_t(r)}{|\dot{r}_p|} \pm \frac{\partial H}{\partial\Linf} \frac{F^\pm_\varphi(r)}{|\dot{r}_p|}
  \right) dr 
\nonumber\\
&
+\int_{r_0}^\infty \frac{H_0}{\sqrt{r-r_0}} 
\left(\mp\frac{\partial r_0}{\partial \Einf}F^\pm_t \pm \frac{\partial r_0}{\partial \Linf}F^\pm_\varphi
\right)dr/|\dot{r}_p| \bigg] .
\end{align}
\end{widetext}
The first four terms here involve double integrals of the self-force. These can be turned into single integrals using integration by parts. For instance, 
\begin{align}\label{ibp0}
& \int_{r_0}^\infty \frac{1}{\sqrt{r-r_0}}\frac{\partial H}{\partial r}\Delta E^\pm(r) dr  =
\nonumber\\
& \left. \left(\int_{r_0}^r \frac{1}{\sqrt{r'-r_0}}\frac{\partial H}{\partial r'} dr' \right)\Delta E^\pm \right|_{r_0}^\infty 
\nonumber\\
& - \int_{r_0}^\infty  \left(\int_{r_0}^r \frac{1}{\sqrt{r'-r_0}}
  \frac{\partial H}{\partial r'}dr' \right)\left(\mp \frac{F^\pm_t}{|\dot{r}_p|}\right)dr 
\nonumber\\
& =   G_r(\infty)\Delta E^\pm(\infty) \pm \int_{r_0}^\infty G_r(r) F^\pm_t\, dr/|\dot{r}_p| ,
\end{align}
where
\begin{equation}\label{Gr}
 G_r(r):= \int_{r_0}^{r}\frac{1}{\sqrt{r'-r_0}}\frac{\partial H(r')}{\partial r'}dr'.
\end{equation}
The lower surface terms in (\ref{ibp0}) vanish: For $r\to r_0$, $\Delta E^\pm$ is bounded (and generally non zero), as is  $\frac{\partial H}{\partial r}$, so the term behaves as $\sim (r-r_0)^{1/2}\to 0$. As for the upper surface term, it too vanishes for the inbound leg, since, for $r\to\infty$, $\Delta E^-(r)\sim rF^-_t\sim 1/r$ at least. However, the upper surface term does not vanish for the outbound leg:
\begin{equation}\label{DeltaE+}
\Delta E^+(\infty)= -\int_{-\infty}^{\infty} F_t \, d\tau = -\sum_{\pm} \int_{r_0}^\infty F_t^\pm dr/|\dot{r}_p|,
\end{equation} 
which describes the total energy radiated.  
Thus, summing over $\pm$ in equation (\ref{ibp0}) we obtain, overall 
\begin{align}\label{ibp0simp_E}
&\sum_{\pm}\int_{r_0}^\infty \frac{1}{\sqrt{r-r_0}}\frac{\partial H}{\partial r}\Delta E^\pm(r) dr
\nonumber\\
&=\sum_{\pm}\int_{r_0}^\infty \big(-G_r(\infty)\pm G_r(r)\big)\, F^\pm_t dr/ |\dot{r}_p|.
\end{align}
Similarly, 
\begin{align}\label{ibp0simp_L}
&\sum_{\pm}\int_{r_0}^\infty \frac{1}{\sqrt{r-r_0}}\frac{\partial H}{\partial r}\Delta L^\pm(r) dr
\nonumber\\
&=\sum_{\pm}\int_{r_0}^\infty \big(G_r(\infty)\mp G_r(r)\big)\, F^\pm_\varphi dr/|\dot{r}_p|,
\end{align}
where we have used
\begin{equation}\label{DeltaL+}
\Delta L^+(\infty)= \int_{-\infty}^{\infty} F_\varphi \, d\tau = \sum_{\pm} \int_{r_0}^\infty F_\varphi^\pm dr/|\dot{r}_p|.
\end{equation} 

We apply a similar integration-by-parts procedure to the first two terms in the second line of (\ref{delta_varphi_v2}). For the first term we thus obtain
\begin{align}\label{ibp2}
&\sum_\pm\int_{r_0}^\infty \sqrt{r-r_0}
  \frac{\partial^2 H}{\partial r\partial\Einf} \Delta E^\pm(r) dr= 
\nonumber\\
&  \sum_{\pm}\int_{r_0}^\infty \left(\frac{1}{2}G_E(\infty) \pm \sqrt{r-r_0}  \frac{\partial H}{\partial\Einf} \mp \frac{1}{2}G_E(r)\right)
  \frac{F_t^\pm}{|\dot{r}_p|}dr,
\end{align}
where 
\begin{equation}\label{GE}
 G_E(r):= \int_{r_0}^{r}\frac{1}{\sqrt{r-r_0}}\frac{\partial H(r')}{\partial\Einf}dr' ,
\end{equation}
and where we have used (\ref{DeltaE+}) again.
Similarly,
\begin{align}\label{ibp2}
& \sum_\pm\int_{r_0}^\infty \sqrt{r-r_0}
  \frac{\partial^2 H}{\partial r\partial\Linf} \Delta L^\pm(r) dr= 
\nonumber\\  
&  \sum_{\pm} \int_{r_0}^\infty \left(-\frac{1}{2}G_L(\infty) \mp \sqrt{r-r_0}  \frac{\partial H}{\partial\Linf} \pm \frac{1}{2}G_L(r)\right)
  \frac{F_\varphi^\pm}{|\dot{r}_p|}dr ,
\end{align}
where
\begin{equation}\label{GL}
 G_L(r):= \int_{r_0}^{r}\frac{1}{\sqrt{r-r_0}}\frac{\partial H(r')}{\partial\Linf}dr' ,
\end{equation}
and where we have used (\ref{DeltaL+}) again. 

With these substitutions, Eq.\ (\ref{delta_varphi_v2}) takes a final form similar to that of (\ref{deltaphi1_final}):
%
\begin{align}\label{deltaphi1_final_method2}
\delta\psi=
\sum_{\pm} \int_{r_0}^{\infty} \left[\tilde{\cal G}^\pm_E(r) F_t^\pm(r)
-\tilde{\cal G}^\pm_L(r) F_\varphi^\pm(r)\right] dr/|\dot{r}_p|,
\end{align}
with
\begin{eqnarray}\label{calG_II}
\tilde{\cal G}^\pm_E(r) &=& \pm G_E(r)-G_E(\infty) 
\nonumber\\
&& +\left(\pm G_r(r) -G_r(\infty) \mp \frac{H(r)}{\sqrt{r-r_0}}\right) \frac{\partial r_0}{\partial\Einf},
\nonumber\\
\tilde{\cal G}^\pm_L(r) &=& \pm G_L(r)-G_L(\infty) 
\nonumber\\
&& +\left(\pm G_r(r) -G_r(\infty) \mp \frac{H(r)}{\sqrt{r-r_0}}\right) \frac{\partial r_0}{\partial\Linf}.
\nonumber\\
\end{eqnarray}
The functions $\tilde{\cal G}^\pm_E(r)$ and $\tilde{\cal G}^\pm_L(r)$ are computed from geodesic relations alone. The final result for $\delta\psi$ in our alternative method, Eq.\ (\ref{deltaphi1_final_method2}), involves a single orbital integral over self-force components. 

One can confirm (and we have done so numerically) that the alternative expression (\ref{deltaphi1_final_method2}) is equivalent to (\ref{deltaphi1_final}). Note, however, that, in general, $\tilde{\cal G}_E(r(\chi)) \ne {\cal G}_E(\chi)$ and 
$\tilde{\cal G}_L(r(\chi)) \ne {\cal G}_L(\chi)$. That is because (\ref{deltaphi1_final_method2}) differs from (\ref{deltaphi1_final}) by surface terms that are only zero if the self-force satisfies certain vanishing conditions at the integration's boundaries. However, the {\it integrals} are equal, assuming the self-force satisfies these conditions.   

We note, finally, that the separate integrals over the $F_t$ and $F_\varphi$ terms in Eq.\ (\ref{deltaphi1_final_method2}) do not individually converge, due to the $\sim (r-r_0)^{-1}$ singularity of the integrands at the periastron; it is only the sum of two terms for which the integral converges. This follows from an analysis similar to that presented in Appendix \ref{App:Convergence}.

\subsection{Conservative and dissipative pieces}

From Eq.\ (\ref{F_cons_diss}) we recall that the conservative piece of the self-force satisfies (for $\alpha=t,\varphi$),
\begin{equation}
F_\alpha^{\rm cons+}(r)= - F_\alpha^{\rm cons-}(r).
\end{equation}
As a result, we can write the conservative contribution as an integral along a single leg of the orbit, as done in Sec \ref{subsec:SFchiCons&Diss}. We obtain
\begin{align}\label{deltaphi1_fcons_method2}
\delta\psi_{\rm cons}=
\int_{r_0}^{\infty} \left(\tilde{\cal G}^{\rm cons}_E F_t^{\rm cons}
-\tilde{\cal G}^{\rm cons}_L F_\varphi^{\rm cons}\right) dr/|\dot{r}_p|,
\end{align}
where 
\begin{align}\label{calG_II_cons}
\tilde{\cal G}^{\rm cons}_E(r) &= 2G_E(r) +2\left(G_r(r) - \frac{H(r)}{\sqrt{r-r_0}}\right) \frac{\partial r_0}{\partial\Einf},
\nonumber\\
\tilde{\cal G}^{\rm cons}_L(r) &= 2G_L(r) +2\left(G_r(r) - \frac{H(r)}{\sqrt{r-r_0}}\right) \frac{\partial r_0}{\partial\Linf}.
\end{align}

Meanwhile, the dissipative components satisfy 
\begin{equation}
F_\alpha^{\rm diss+}(r)=  F_\alpha^{\rm diss-}(r),
\end{equation}
(for $\alpha=t,\varphi$)
from which we obtain 
\begin{align}\label{deltaphi1_fdiss_method2}
\delta\psi_{\rm diss}=
\int_{r_0}^{\infty} \left(\beta_E F_t^{\rm diss}
-\beta_L F_\varphi^{\rm diss}\right) dr/|\dot{r}_p|,
\end{align}
where $\beta_E,\beta_L$ are {\it constants} given by 
\begin{align}\label{beta}
\beta_E &= -2\left(G_E(\infty)+G_r(\infty)\frac{\partial r_0}{\partial\Einf} \right), 
\nonumber\\
\beta_L& = -2\left(G_L(\infty)+G_r(\infty)\frac{\partial r_0}{\partial\Linf} \right).
\end{align}
It can be checked that 
\begin{equation}\label{beta_alpha}
\beta_E = -\alpha_E \Einf, \quad\quad  \beta_L = -\alpha_L \Linf, 
\end{equation}
confirming that (\ref{deltaphi1_fdiss_method2}) is equivalent to (\ref{deltavarphi_diss}). In terms of the $\beta$ coefficients, Eq.\ (\ref{deltavarphi_diss_fluxes}) becomes
\begin{equation}
\delta\psi_{\rm diss} = -\frac{1}{2}\left(\beta_E E_{\rm rad} + \beta_L L_{\rm rad}\right).
\end{equation}

Equations (\ref{deltaphi1_final_method2}), (\ref{deltaphi1_fcons_method2}) and (\ref{deltaphi1_fdiss_method2}) constitute the final results, in our alternative formulation, for, respectively, the full self-force correction $\delta\psi$, its conservative piece and its dissipative piece. We will implement these formulas numerically in Secs.\ \ref{Sec:ScalarImplementation} and \ref{Sec:results}.

\section{Weak-field limit}
\label{sec:PMlimit}

It is instructive to extract the weak-field limit of our formulas for $\delta\psi$, not least for the purpose of checking our expressions against known PM results. In this section we derive the leading-order PM reduction of the expressions derived above for $\delta\psi_{\rm cons}$ and $\delta\psi_{\rm diss}$. First we do so without PM-expanding the self-force components themselves, leading to simple weak-field formulas for the conservative and dissipative pieces of $\delta\psi$ in terms of integrals over Cartesian components of the self-force. Then, we substitute the leading-order PM self-force derived by Gralla and Lobo in \cite{GrallaLobo2022}, and verify that our result for $\delta\psi$ agrees with theirs. This provides an overall check on the validity of the general expressions derived for $\delta\psi$ in previous sections. 

Let us thus consider the PM reduction of Eqs.\ (\ref{deltaphi_methodI_cons}) and (\ref{deltavarphi_diss}) for $\delta\psi_{\rm cons}$ and $\delta\psi_{\rm diss}$ [we have checked that the equivalent formulas (\ref{deltaphi1_fcons_method2}) and (\ref{deltaphi1_fdiss_method2}) yield the same leading-order PM reductions, as expected].
Substituting the geodesic PM expansions from Sec.\ \ref{subsec:PM} in Eqs.\ (\ref{tauchi}), (\ref{alphaE}), (\ref{alphaL}) and (\ref{calF}), and re-expanding in powers of $M/b$ at fixed $v$, we obtain 
\begin{equation}\label{tauchi_PM}
\tau_\chi = \frac{b}{v E}\sec^2\chi  +O(b^0),
\end{equation}
and
\begin{eqnarray}\label{alphaEalphaL_PM}
\alpha_E &=& -\frac{2 M \left(1-3 v^2\right)}{b E^2 v^4} + O(b^{-2}), \nonumber\\
\alpha_L &=& -\frac{2 M \left(1+v^2\right)}{b^3 E^2 v^4} + O(b^{-4}),
\end{eqnarray}
as well as
\begin{eqnarray}
{\cal F}_E = \frac{\cot (\chi )}{E v^2} + O(b^{-1}), \quad\quad
{\cal F}_L = -\frac{\cot (\chi )}{b E v} + O(b^{-2}). \nonumber
\end{eqnarray}
In turn, substituting these leading-order expressions in Eqs.\ (\ref{deltaphi_methodI_cons}) and (\ref{deltavarphi_diss}), and recalling Eq.\ (\ref{chiinfPM}), we arrive at 
\begin{equation}\label{1PM_methodI}
\delta\psi_{\rm cons}\overset{\rm PM}{\sim} \frac{4}{v^2E^2}\int_0^{\pi/2}\left[(b/v)F^{\rm cons}_t+F^{\rm cons}_{\varphi}\right]\frac{d\chi}{\sin 2\chi},
\end{equation}
and
\begin{align}
\label{1PM_methodIdiss}
\delta\psi_{\rm diss}\overset{\rm PM}{\sim}  -\frac{2 M}{bE^2 v^4} \int_0^{\pi/2}\Big[&(b/v)\left(3 v^2-1\right) F^{\rm diss}_t
\nonumber\\
&+\left(1+v^2\right) F^{\rm diss}_{\varphi }\Big] \, \sec^2\!\chi \, d\chi,
\end{align}
where $\overset{\rm PM}{\sim}$ denotes equality at leading PM order. 
It is useful to re-express these results in terms of radial integrals, which we can do with the help of the leading-order relation 
\begin{equation}
\cos\chi \overset{\rm PM}{\sim} \frac{b}{r},
\end{equation}
itself derived by substituting (\ref{ePM}) and (\ref{pPM}) in (\ref{rofchi}). We obtain 
\begin{equation}\label{1PM_methodII_rad}
\delta\psi_{\rm cons} \overset{\rm PM}{\sim} \frac{2}{v^2E^2}\int_{b}^{\infty}\left[(b/v)F_t^{\rm cons}+F_{\varphi}^{\rm cons}\right] \frac{r\, dr}{r^2-b^2},
\end{equation}
\begin{align}
\label{1PM_methodIIdiss_rad}
\delta\psi_{\rm diss} \overset{\rm PM}{\sim}  -\frac{2 M}{b^2E^2 v^4} \int_b^{\infty}\big[&(b/v)\left(3 v^2-1\right) F^{\rm diss}_t \nonumber\\
&+\left(1+v^2\right) F^{\rm diss}_{\varphi }\big]\, \frac{r\, dr}{\sqrt{r^2-b^2}}\, .
\end{align}
While we have stopped at leading order, it should be straightforward to derive higher-order terms in the PM expansions of $\delta\psi_{\rm cons}$ and $\delta\psi_{\rm diss}$.

In the weak-field limit, the scattering orbit is more naturally described in Cartesian (rather than polar) coordinates. Let us introduce (following Ref.\ \cite{GrallaLobo2022}) a Cartesian coordinate system $(t,x,y,z)$ centered at the large mass $M$, such that, in the weak-field limit, the scattering orbit approaches the straight line $x_p^\mu(t)=(t,b,0,z(t))$, where $z=vt=\pm\sqrt{r^2-b^2}$. The moment $t=0$ corresponds to the point of closest approach, where $r=b$ and $z=0$. Using $\frac{\partial\varphi}{\partial x}=-\frac{z}{r^2}$ and $\frac{\partial\varphi}{\partial z}=\frac{x}{r^2}$ we then have
\begin{equation}
F_t = -v F^z,
\quad\quad
F_\varphi = -z F^x+b F^z ,
\end{equation}
and Eqs.\ (\ref{1PM_methodII_rad}) and (\ref{1PM_methodIIdiss_rad}) become 
\begin{equation}\label{1PM_methodII_Cart}
\delta\psi_{\rm cons} \overset{\rm PM}{\sim} - \frac{2}{v^2E^2}\int_{0}^{\infty}F^x_{\rm cons} dz,
\end{equation}
\begin{align}
\label{1PM_methodIIdiss_Cart}
\delta\psi_{\rm diss} \overset{\rm PM}{\sim}  -\frac{2 M}{b^2E^2 v^4} \int_0^{\infty}\big[&2b\left(1-v^2\right) F_{\rm diss}^z \nonumber\\
&-\left(1+v^2\right) z F_{\rm diss}^x\big]dz ,
\end{align}
where we have also used $r^2-b^2=z^2$. 


Ref.\ \cite{GrallaLobo2022} provides analytical expressions for the full (dissipative+conservative) gravitational self-force, in the $M$-centered system, at leading PM order. The force can be written in the form 
\begin{equation}\label{F_PM}
F^\alpha = \frac{\hat F^\alpha(\hat z; v)}{b^3},
\end{equation}
where $\hat z:= z/b $, and $\hat F^\alpha$ depends only on $\hat z$ (as a dimensionless parameter along the orbit) and on $v$, but not otherwise on $b$. The conservative and dissipative pieces of $F^\alpha$ can then be extracted using
$F^\alpha_{\rm cons}(z) = \frac{1}{2}\left[F^\alpha(z) \pm F^\alpha(-z) \right]$ and 
$F^\alpha_{\rm diss}(z) = \frac{1}{2}\left[F^\alpha(z) \mp F^\alpha(-z) \right]$, 
with the upper sign for $\alpha=x$ and the lower sign for $\alpha=z$. Substituting in (\ref{1PM_methodII_Cart}) and (\ref{1PM_methodIIdiss_Cart}) and changing the integration variable from $z$ to $\hat z$, we immediately see that
\begin{equation}
\delta\psi_{\rm cons} \overset{\rm PM}{\sim} O{(M/b)^2},
\quad\quad
\delta\psi_{\rm diss} \overset{\rm PM}{\sim} O{(M/b)^3},
\end{equation}
i.e., the leading conservative and dissipative self-force contributions to the scattering angle occur, respectively, at 2PM and 3PM orders, as expected. 

The explicit expressions for $\hat F^\alpha(\hat z; v)$ are rather lengthy, and can be found in Section 4.1 of \cite{GrallaLobo2022}. [To convert to our notion, identify $f^z_{(m)}$ and $f^x_{(m)}$ in their Eqs.\ (37) and (38) with our $\hat F^z$ and $\hat F^x$, respectively; replace in these equations $m\to 1$ and $\gamma\to E$; and in their Eqs.\ (39)--(47) replace $z\to\hat z$ and $r\to r/b=\sqrt{1+\hat z^2}$.] Despite the unwieldiness of the explicit expressions for $F^\alpha_{\rm cons}$ and $F^\alpha_{\rm diss}$, the $\hat z$ integrals in Eqs.\  (\ref{1PM_methodII_Cart}) and (\ref{1PM_methodIIdiss_Cart}) are elementary, and yield the simple final results
\begin{align}\label{deltapsi_cons_PM}
\delta\psi_{\rm cons} \overset{\rm PM}{\sim}  & \:\; \frac{7\pi }{4}\left(\frac{M}{b}\right)^2 , \\
\delta\psi_{\rm diss} \overset{\rm PM}{\sim} & -\frac{22 E}{3}\frac{(1+v^2)^2}{v^3}\left(\frac{M}{b}\right)^3 .\label{deltapsi_diss_PM}
\end{align}

This result for $\delta\psi_{\rm cons}$ agrees with that obtained in \cite{GrallaLobo2022} [see Eqs.\ (131) with (128) therein\footnote{Note there is an overall factor $v^2$ missing in Eq.\ (128) of \cite{GrallaLobo2022}, due to a misprint \cite{priv_comm_Lobo}.}] using a different method. Ref.\ \cite{GrallaLobo2022} neglects 3PM terms of the scattering angle, so a similar comparison is not possible for $\delta\psi_{\rm diss}$. It is important to note that, in the gravitational self-force problem, our $\delta\psi$ differs from the ``physical'' scattering angle commonly considered in the literature (and usually denoted by $\chi$), in that (1) $\delta\psi$ is calculated in a (noninertial) $M$-centered system rather than in a center-of-mass system, and (2) $\delta\psi$ neglects the contribution from the so-called ``matter-dominated force'' discussed in \cite{GrallaLobo2022}. To relate $\delta\psi_{\rm diss}$ in Eq.\ (\ref{deltapsi_diss_PM}) to the known 3PM dissipative term of $\chi$ [see e.g.\ Eq.\ (7.36) of \cite{Herrmann2021}] would require a calculation of both corrections to 3PM order, which we do not attempt here.

\section{Scalar-charge model}
\label{Sec:Scalar}

We wish to illustrate the numerical implementation of Eqs.\ (\ref{deltaphi_methodI_cons}) and (\ref{deltavarphi_diss}) [or (\ref{deltaphi1_fcons_method2}) and (\ref{deltaphi1_fdiss_method2})] with the full self-force. However, numerical results for the gravitational self-force along a scattering orbit are not yet available (cf.\ \cite{LongBarack2021} for a discussion of progress and prospects). Instead, we content ourselves here with a numerical illustration based on a scalar-charge toy model, for which numerical results may be obtained with relative ease using the time-domain method developed in Ref.\ \cite{LongBarack2021}. The structure of the self-forced equations of motion, and thus also the form of Eqs.\ (\ref{deltaphi_methodI_cons}) and (\ref{deltavarphi_diss}) [or (\ref{deltaphi1_fcons_method2}) and (\ref{deltaphi1_fdiss_method2})] for the scattering angle, carry over almost intact from the gravitational problem to the scalar one, which makes our toy model particularly instructive. At the same time, the scalar model completely avoids the gauge-adjustment issues alluded to at the end of the previous section. The issue of gauge will need to be tackled separately for the gravitational problem.

\subsection{Equation of motion}

In the scalar-charge toy model we endow the particle with a scalar charge $Q$, assume $Q\ll \sqrt{\mu M}$, and ignore the gravitational self-force. The scalar charge sources a scalar field $\Phi\propto Q$, which, we assume, is massless, minimally coupled, and satisfies the Klein-Gordon equation on the Schwarzschild background, 
\begin{equation}
\nabla^\alpha\nabla_\alpha \Phi = -4\pi Q \int_{-\infty}^{\infty} \frac{\delta^4(x-x_p(\tau))}{\sqrt{-g(x)}}d\tau.
\label{eqn:Scalar3+1}
\end{equation}
Here $x_p(\tau)$ again denotes the particle's orbit, $\tau$ is its proper time, and $g$ is the determinant of the Schwarzschild metric. 
Back-reaction from the scalar field exerts on the particle a self-force $\propto Q^2$. The equation of motion, analogous to (\ref{EoM_SF}), is
\begin{equation}\label{eom_scalar}
u^\beta \nabla_\beta(\mu u^\alpha) = Q\nabla^\alpha \Phi^R =: {\cal F}^\alpha,
\end{equation}
where $u^\beta$ is the four-velocity, and $\Phi^R$ is the Detweiler-Whiting regular piece of $\Phi$ \cite{DetweilerWhiting2003}, here evaluated at the particle, $x=x_p$. The quantity ${\cal F}^\alpha(\propto Q^2)$ is the self-force due to the scalar field. A method for constructing $\Phi^R$ and ${\cal F}^\alpha$ in practice will be reviewed in Sec.\ \ref{subsec:mode-sum} below.  

It is useful to split Eq.\ (\ref{eom_scalar}) into its orthogonal-to-$u^\alpha$ and tangent-to-$u^\alpha$ components:
\begin{eqnarray}
u^\beta \nabla_\beta u^\alpha &=& (\delta_{\beta}^{\alpha}+u^\alpha u_\beta){\cal F}^\beta/\mu =: q_s F^\alpha, \label{orthogonal}
\\
\frac{d\mu}{d\tau} &=& -u^\alpha {\cal F}_\alpha =: -{\cal F}_u . \label{parallel}
\end{eqnarray}
Here we have introduced
\begin{equation}
q_s: = \frac{Q^2}{\mu M}\ll 1, 
\end{equation}
which plays the role of the small mass ratio $q$ in the gravitational problem [compare the form of Eq.\ (\ref{orthogonal}) to that of (\ref{EoM_SF})].
The orthogonal self-force component in Eq.\ (\ref{orthogonal}) gives rise to self-acceleration; it is analogous to the gravitational self-force $F^\alpha$ featuring in previous sections, and we adopt the same notation for both quantities in order to later enable us to reuse some of our scattering-angle expressions (this should cause no confusion, since in the rest of this work we discuss the scalar-field model exclusively). The tangent component of the self-force in Eq.\ (\ref{parallel}), ${\cal F}_u$, causes the rest mass $\mu$ of the particle to evolve, trading energy-mass between the particle and the scalar field. (This contrasts with the situation in the gravitational problem, where ${\cal F}_u$ vanishes identically and the rest mass is conserved.) In fact, after substituting for $\cal F_\alpha$ from Eq.\ (\ref{eom_scalar}),  Eq.\ (\ref{parallel}) can be immediately integrated to give 
\begin{equation}
\mu(\tau) = \mu_0 - Q \Phi^R(\tau),
\label{eqn:ScalarMass}
\end{equation}
where $\mu_0$ is a constant of integration. In the scattering scenario we expect $\Phi^R(-\infty)=\Phi^R(+\infty)$, so there should be no net change in rest mass overall.  

\subsection{Scattering angle}

The self-force equation of motion (\ref{orthogonal}) has the same form as the analogous gravity-case equation (\ref{EoM_SF}), with the simple replacement $q\to q_s$. The solutions to the equation of motion will also take the same form, simply replacing the gravitational self-force with the (orthogonal component of the) scalar-field one, and the mass ratio $q$ with $q_s$. In particular, with these identifications, the formulas derived in Secs.\ \ref{Sec:SFchi} and \ref{Sec:SFr} for the self-force-corrected scattering angle apply also in the scalar-charge model. Specifically, if we write the perturbed scattering angle in the form
\begin{equation}\label{phisplit_scalar}
\tilde\psi = \psi + q_s\, \delta\psi,
\end{equation}
where the split between background $\psi$ and perturbation $q_s\, \delta\psi$ is, as always, defined with fixed ($v,b$), then the conservative and dissipative pieces of $\delta\psi$ are still given by Eqs.\ (\ref{deltaphi_methodI_cons}) and (\ref{deltavarphi_diss}) [or (\ref{deltaphi1_fcons_method2}) and (\ref{deltaphi1_fdiss_method2})], now with $F^\alpha$ identified as the (orthogonal component of the) scalar-field self-force. The variation of rest mass $\mu$ due to ${\cal F}_u$ does not affect this conclusion, since in our model $\mu$ is taken to be solely inertial and does not self-gravitate. In deriving Eqs.\ (\ref{deltaphi_methodI_cons}) and (\ref{deltavarphi_diss}) [or (\ref{deltaphi1_fcons_method2}) and (\ref{deltaphi1_fdiss_method2})] we have assumed that $F^\alpha$ falls off sufficiently fast at infinity for various integration surface term to vanish; it can be checked that these assumptions remain true of the scalar-field self-force as well, and we have confirmed that with our numerical calculation.

\subsection{Weak-field limit}

Ref.\ \cite{GrallaLobo2022} has derived the leading-order PM term of the scalar-field self-force as well, together with the associated correction to the scattering angle. We can use these results to perform additional checks on the validity of our expressions for $\delta\psi$. 

Equations (33) and (34) of \cite{GrallaLobo2022} give the Cartesian components of the full (conservative+dissipative) scalar-field self-force in the frame of the large mass $M$. Once again they take the form shown in Eq.\ (\ref{F_PM}) above. We can again extract the conservative and dissipative components of the force using the $z\to -z$ symmetry as in the gravitational case. Substituting the results (which are again rather unwieldy) in the leading-order PM formulas (\ref{1PM_methodII_Cart}) and (\ref{1PM_methodIIdiss_Cart}), we obtain, for the scalar-charge model,
\begin{align}\label{PM_cons_scalar}
\delta\psi_{\rm cons} \overset{\rm PM}{\sim} &  -\frac{\pi}{4}\left(\frac{M}{b}\right)^2, \\
\delta\psi_{\rm diss} \overset{\rm PM}{\sim} & \:\;  \frac{2E}{3}\frac{(1+v^2)^2}{v^3}\left(\frac{M}{b}\right)^3 . \label{PM_diss_scalar}
\end{align}
The expression for $\delta\psi_{\rm cons}$ agrees with that derived in \cite{GrallaLobo2022} [see Eq.\ (2) there] using a different method. Additionally, both expressions agree with calculations performed using quantum scattering amplitude techniques \cite{priv_comm_Amp}.

\subsection{Construction of $\Phi^R$ and of ${\cal F}^\alpha$}
\label{subsec:mode-sum}

Our numerical calculation of $\Phi^R$ and of ${\cal F}^\alpha$ along scattering orbits, in later sections,  will be based on the standard method of mode-sum regularization \cite{BarackOri2000,Barack2009}. We review here the basic prescription.  

Consider the decomposition of the scalar field $\Phi$ into a basis of spherical harmonics $Y_{\ell m} (\theta, \varphi)$ defined on spheres $t,r={\rm const}$ around the large black hole: 
\begin{equation}
\Phi = \frac{2\pi Q}{r} \sum_{\ell =0}^{\infty} \sum_{m=-\ell}^{\ell} \phi_{\ell m}(t,r) Y_{\ell m} (\theta, \varphi).
\end{equation}
Similarly decomposing the source side of Eq.\ (\ref{eqn:Scalar3+1}), we derive decoupled modal equations for the (complex) time-radial fields $\phi_{\ell m}(t,r)$:
\begin{align}\label{eqn:SourcedFieldEquation}
\frac{\partial^2\phi_{\ell m}}{\partial t^2} - &\frac{\partial^2\phi_{\ell m}}{\partial r_*^2}  + U(r) \phi_{\ell m} 
\nonumber\\
& =\frac{2 f(r_p)^2}{E r_p(t)} \delta \left(r - r_p(t)\right) \bar{Y}_{\ell m}(\theta_p,\varphi_p(t)).
\end{align}
where $r_* = r+2M \log[r/(2M)-1]$ is the standard Schwarzschild tortoise coordinate, an overbar denotes complex conjugation, and
\begin{equation}
U(r) := f(r)\left(\frac{\ell(\ell+1)}{r^2} + \frac{2M}{r^3} \right).
\label{eq:ScalarPotential}
\end{equation}
We recall our notation: $f=1-2M/r$, and $x^\alpha_p(t)$ describes the scattering trajectory (here using $t$ rather than $\tau$ as parameter along the orbit), with $\theta_p\equiv \pi/2$. 
We take $\Phi$ to be the {\em retarded} solution of the Klein-Gordon equation (\ref{eqn:Scalar3+1}), i.e., the (unique) regular solution that contains no incoming radiation at past null infinity and no outgoing radiation through the past event horizon. Corresponding boundary conditions are imposed on the modal fields $\phi_{\ell m}(t,r)$. 

While the full retarded solution $\Phi$ has the usual Coulomb-like divergence at the particle, the fields $\phi_{\ell m}$ are each finite and continuous at $r=r_p(t)$. Their derivatives remain finite but are generally discontinuous on the particle. The total $\ell$-mode contribution to the value of the scalar field along the particle's worldline,
\begin{equation}
\phi_\ell(t) := \frac{2\pi Q}{r_p(t)} \sum_{m=-\ell}^\ell  \phi_{\ell m}(t,r_p(t)) Y_{\ell m} (\pi/2, \varphi_p(t)),
\end{equation}
is thus finite and well defined, although the sum over $\ell$ modes diverges. In the mode-sum formulation, the Detweiler-Whiting regular piece of the scalar field, $\Phi^R$ [the piece that features in the equation of motion (\ref{eom_scalar})], is constructed as a function along the orbit via the mode sum
\begin{equation}\label{ModeSum_Phi}
\Phi^R(t) = \sum_{\ell=0}^\infty \left[ \phi_\ell(t) - B(t) \right],
\end{equation}
with the ``regularization parameter''
\begin{equation}
B(t)= \frac{2 Q {\cal K}}{\pi \sqrt{L^2+r_p^2}}.
\end{equation}
Here we introduced
\begin{equation}
{\cal K} := \El_1 \left(\frac{\pi}{2}; \frac{L^2}{L^2+r_p^2}\right), \nonumber\\
\end{equation}
where, recall, $\El_1$ is the incomplete elliptic integrals of the first kind [cf.\ Eqs.\ (\ref{eqn:El1})]. The summand in Eq.\ (\ref{ModeSum_Phi}) falls off at large $\ell$ at least as $\ell^{-2}$, so the sum converges at least as $1/\ell$. 

To compute the self-force ${\cal F}^\alpha$, we first construct the modal derivatives
\begin{equation}
{\cal F}_\alpha^{\ell\pm}(t) := \lim_{x \to x_p^{\pm}(t)} \nabla_\alpha \left[\frac{2\pi Q^2}{r}\sum_{m=-\ell}^\ell  \phi_{\ell m}(t,r) Y_{\ell m} (\theta, \varphi)\right]
\end{equation}
as functions along the orbit. Here the $\pm$ refers to whether the limit to the particle is taken from $r\to r_p^+(t)$ or from $r\to r_p^-(t)$, which generally yields two different values. The quantities ${\cal F}_\alpha^{\ell\pm}$ are finite, and grow linearly with $\ell$ at large $\ell$.
In the mode-sum method, the physical self-force along the orbit is then constructed using the mode-sum formula
\begin{equation}
{\cal F_\alpha}(t) = \sum_{\ell=0}^\infty \left[ {\cal F}^{\ell\pm}_\alpha(t) - A^{\pm}_\alpha(t) (\ell+1/2) - B_\alpha(t) \right],
\label{eqn:ModeSum}
\end{equation}
where the nonvanishing components of the regularisation parameters are \cite{regpar}
\begin{eqnarray}
A_t^\pm &=& \frac{\pm Q^2 \dot r_p }{(L^2+r_p^2)}, \nonumber \\
A_r^\pm &=& -\frac{\pm Q^2E}{f_p(L^2+r_p^2)}, \nonumber \\
B_t &=& -\frac{Q^2 E r_p \dot r_p}{\pi (L^2+r_p^2)^{3/2}} (2{\cal E - K}), \nonumber \\
B_r &=& \frac{\left(2E^2r_p^2-f_p(L^2+r_p^2)\right)Q^2{\cal E}}{f_p r_p \pi \left(L^2+r_p^2\right)^{3/2}}\nonumber \\
&&-\frac{\left(E^2r_p^2+f_p(L^2+r_p^2)\right)Q^2{\cal K}}{f_p r_p \pi \left(L^2+r_p^2\right)^{3/2}},\nonumber \\
B_\varphi &=& -\frac{Q^2 r_p \dot r_p}{L\pi\sqrt{L^2+r_p^2}} ({\cal E} - {\cal K}),
\label{eqn:RegParams}
\end{eqnarray}
Here $f_p:=1-2M/r_p$, an overdot denotes $d/d\tau$, and we have introduced
\begin{equation}
{\cal E} := \El_2 \left(\frac{\pi}{2}; \frac{L^2}{L^2+r_p^2}\right),
\end{equation}
recalling that $\El_2$ is the incomplete elliptic integrals of the second kind [cf.\ Eq.\ (\ref{eqn:El2})]. The summand in Eq.\ (\ref{eqn:ModeSum}) no longer depends on the direction in which the limit to the particle is taken. It falls off at large $\ell$ at least as $\ell^{-2}$, so this mode sum too converges at least as $1/\ell$. Once ${\cal F}_\alpha^{\ell\pm}$ has been obtained, its orthogonal and tangent components can be extracted by applying the suitable projection operator, as in Eqs.\ (\ref{orthogonal}) and (\ref{parallel}), to obtain $F^\alpha$ and ${\cal F}_u$ as functions along the orbit.

The large-$\ell$ behavior of the summands in both Eqs.\ (\ref{ModeSum_Phi}) and (\ref{eqn:ModeSum}) has been derived analytically in terms of an expansion in powers of $1/\ell$ \cite{Heffernan2012}, and this can be used to improve the convergence of the mode sums. In our calculation we will make use of terms up to $O(\ell^{-6})$ for that purpose. Expressions for the high-order expansion terms can be found in \cite{Heffernan2012} or within the \texttt{RegularizationParameters} package of the Black Hole Perturbation Toolkit \cite{BHPToolkit}.

\section{Numerical implementation: Method}
\label{Sec:ScalarImplementation}

In this section we review our method for numerically calculating the scalar-field self-force correction to the scattering angle; a sample of results will be presented and discussed in the next section.

Our method is based on a numerical evolution of the modal scalar-field equation (\ref{eqn:SourcedFieldEquation}) in the time domain, from characteristic initial data, using a finite-difference scheme in characteristic coordinates. Our code is a simple adaptation of the code we used in \cite{LongBarack2021} to evolve the Regge-Wheeler equation for a massive particle on a scattering orbit. We thus only give here a general description of our code, referring the reader to Ref.\ \cite{LongBarack2021} for full details. The code takes as input the parameters of a geodesic scattering orbit (along with a range of numerical control parameters; see below), and returns the retarded-field modes $\phi_{\ell m}$ and their derivatives along the scattering orbit. From these we numerically construct the regular field $\Phi^R$ and self force ${\cal F}^\alpha$ (as functions along the orbit) using mode-sum regularization. The conservative and dissipative pieces of the scattering angle are then computed by numerically evaluating the orbital integrals in Eqs.\ (\ref{deltaphi_methodI_cons}) and (\ref{deltavarphi_diss}), and then again, as a check, also using Eqs.\ (\ref{deltaphi1_fcons_method2}) and (\ref{deltaphi1_fdiss_method2}).

\subsection{Numerical algorithm}
\label{Subsec:ScalarMethod}

\begin{figure}[h!]
\centering
\includegraphics[width=\linewidth]{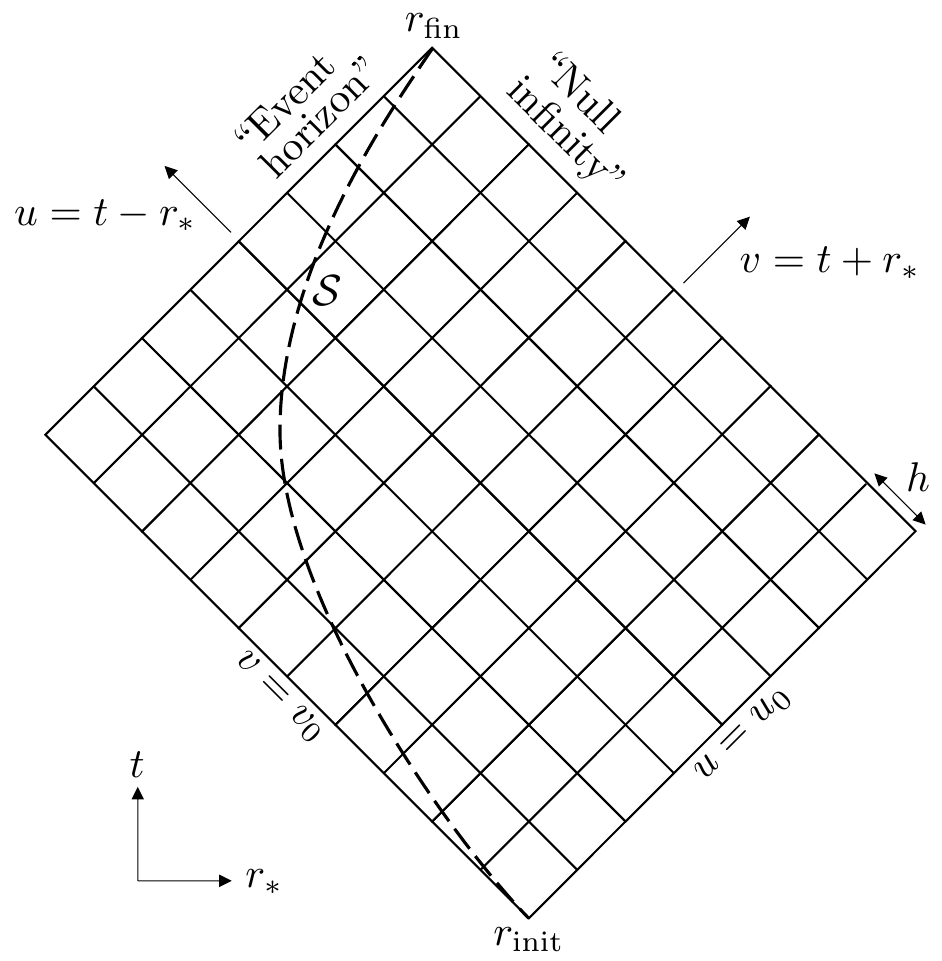}
\caption{Illustration of the 1+1D characteristic grid used in our numerical evolution of the scalar-field modes $\phi_{\ell m}(t,r)$ outside a Schwarzschild black hole. Grid cells have uniform dimensions $h\times h$ in Eddington-Finkelstein coordinates $u,v$. The grid is constructed such that the particle enters and exits at suitable radii $r_{\rm init}$ and $r_{\rm fin}(<r_{\rm init})$, respectively; the inbound leg is extended to the past to enable dissipation of junk radiation. We set initial conditions along the rays $u=u_0$ and $v=v_0$. The dashed line $\cal S$ represents the scalar charge's scattering worldline, which is fixed in advance of the evolution given the geodesic parameters.  The evolution proceeds along successive $u=$ const rays, using the quadratically convergent finite-difference scheme described in Appendix \ref{app:FDS}.}
\label{uvGrid}
\end{figure}

The numerical evolution of Eq.\ (\ref{eqn:SourcedFieldEquation}) is carried out on a fixed mesh in Eddington-Finkelstein coordinates $v=t+r_*$ and $u=t-r_*$, as depicted in Fig.\ \ref{uvGrid}. We use a uniform grid spacing $\Delta v=h=\Delta u$, where, in our runs for this work, $h$ is typically taken in the range $[M/100,M/500]$. The two initial rays $u=u_0$ and $v=v_0$ are chosen such that the wordline representing the scattering geodesic orbit (denoted $\cal S$ in the figure)  intersects the initial vertex $(u_0,v_0)$. We supply characteristic initial data (see below) on the initial rays and then evolve the data using a finite-difference version of Eq.\ (\ref{eqn:SourcedFieldEquation}), detailed in Appendix \ref{app:FDS}. The finite-difference scheme has a local discretization error of $O(h^4)$ [or $O(h^3)$ for grid cells intersected by the particle's worldline], which ensures that the global accumulated error in the field scales as $h^2$. We have tested and confirmed the quadratic convergence of our code by comparing results obtained with a sequence of decreasing $h$ values.

For characteristic initial conditions we simply set $\phi(u_0,v)\equiv 0$ and $\phi(u,v_0)\equiv 0$. The unphysical data produces an outburst of spurious (`junk') radiation, which, however, decays at late time (theoretically, as $t^{-2\ell-3}$ \cite{Barack:1998bw}). Later we discard the junk-contaminated portion of the data. To determine what portion of the data is sufficiently junk-free, we run with different initial radii $r_{\rm init}$ and compare; see Fig.\ 8 of \cite{LongBarack2021} and the discussion around it. Note that our characteristic numerical domain has no timelike boundaries, so no boundary conditions need be imposed. 

The initial and final radii, $r_{\rm init}$ and $r_{\rm fin}$, are input parameter in our code, and are set so that a sufficiently long segment of clean data remains after the removal of the junk-contaminated portion. We choose $r_{\rm init}$ such that the junk has sufficiently radiated away before the particles reaches $r=r_{\rm fin}$ on the {\em inbound} leg of the orbit, so that we obtain clean data over the entire range $r_0\leq r\leq r_{\rm fin}$ on both inbound and outbound legs.  Our rough criterion in choosing $r_{\rm fin}$ is that the error in the final scattering angle due to the large-$r$ truncation of the orbit remained subdominant compared to other sources of numerical error (see our discussion of error estimate below), except in cases where this would take us beyond the limits of the computational resources available to us. This criterion meant that $r_{\rm fin}$ had to be adjusted as a function of the orbital parameters: weak-field orbits have larger relative contribution to the scattering angle coming from larger radii, so they require a larger $r_{\rm fin}$. For the data presented in this work we have used $r_{\rm fin}$ values between $200M$ (strongest-field orbit; required $r_{\rm init}\sim 260M$) and $600M$ (weakest-field orbit; required $r_{\rm init}\sim 2000M$ ). The evolution terminates when the scattered particle reaches $r=r_{\rm fin}$ on the outbound leg.  Note that the run time of our 1+1D evolution code scales roughly {\it quadratically} with $r_{\rm fin}$, so increasing $r_{\rm fin}$ is strongly punitive computationally. 

Below we lay out the main steps of our numerical algorithm.

{\it Input.} The code takes as input the two orbital parameters $v$ and $b$, the initial and final orbital radii $r_{\rm init}$ and $r_{\rm fin}$, the maximum multipole number $\ell_{\rm max}$, and the finite-difference interval $h$. 

{\it Step 1: Calculate geodesic orbit.} Given $v$ and $b$, the code calculates $E$ and $L$ and from these $e$ and $p$, as well as the periastron distance $r_0$. The functions $r_p(t)$ and $\varphi_p(t)$ are then computed in the range $r_0\leq r_p\leq r_{\rm init}$, by numerically integrating Eqs.\ (\ref{tdot})--(\ref{rdot}) with the initial conditions $r_p(0)=r_0$ and $\varphi_p(-\infty)=0$. The code also calculates $t_{\rm init}$ and $t_{\rm fin}$, the values of $t$ associated with $r_{\rm init}$ (on the inbound leg) and $r_{\rm fin}$ (on the outbound leg), respectively.

{\it Step 2: Set characteristic grid.} The code then prepares a $2\times 2$ array of $u,v$ coordinate values representing the nodes of the characteristic mesh shown in Fig.\ \ref{uvGrid}. For the initial rays we take $u_0=t_{\rm init}-r^*_{\rm init}$ and $v_0=t_{\rm init}+r^*_{\rm init}$ with $r^*_{\rm init}:=r_*(r_{\rm init})$. This is so that the initial vertex $(u,v)=(u_0,v_0)$ is crossed by the particle at $(t,r)=(t_{\rm init},r_{\rm init})$. The stepping interval is set at $h$, and the grid's dimensions are taken such that the apex cell contains the point $(t,r)=(t_{\rm fin},r_{\rm fin})$. The particle then exits the numerical domain at a radius $r^*_{\rm fin} := r_*(r_{\rm fin})+\epsilon$, with $\epsilon \leq h/2$, on its way out (the exit point need not generally correspond to a grid point). Finally, the coordinate values of all intersections of the orbit with grid lines are calculated and stored. 

{\it Step 3: Obtain the scalar modes $\phi_{\ell m}$.} We evolve the 1+1D scalar field equation (\ref{eqn:SourcedFieldEquation}) using the second-order-convergent finite-difference scheme described in Appendix \ref{app:FDS} for each $-\ell\leq m\leq \ell$ with $0\leq \ell\leq\ell_{\rm max}$. The scheme requires as input the values of $r_p(t)$ and $\varphi_p(t)$ (as well as their first derivatives) at intersections of the worldline with grid lines. The evolution starts with zero initial data along $v=v_0$ and $u=u_0$ and proceeds along successive lines of $u=$ const. The outcome is a finite-difference approximation to the modes $\phi_{\ell m}$.

{\it Step 4: Mode-sum regularisation.} We use the results of the previous step to calculate the field modes along the orbit. The code then applies the mode-sum formulae (\ref{ModeSum_Phi}) and (\ref{eqn:ModeSum}) to compute $\Phi^R$ and ${\cal F}^\alpha$ at each intersection of the worldline with grid lines, and from the latter we construct $F_t^{\rm cons},F_\varphi^{\rm cons},F_t^{\rm diss}$ and $F_\varphi^{\rm diss}$ using Eqs.\ (\ref{orthogonal}) and (\ref{F_cons_diss}). For the mode sums we take $\ell_{\rm max}=15$, and incorporate high-order regularisation parameters down to order $\ell^{-6}$.

{\it Step 5: Calculate correction to the scattering angle.}  The self-force datasets from the previous step are then interpolated along the orbit, and the interpolations are used to form the integrands in Eqs.\ (\ref{deltaphi_methodI_cons}) and (\ref{deltavarphi_diss}), as well as  (\ref{deltaphi1_fcons_method2}) and (\ref{deltaphi1_fdiss_method2}). Finally, we integrate numerically in these equations to obtain $\delta\psi_{\rm cons}$ and $\delta\psi_{\rm diss}$. The integration error is estimated by comparing the results of the $\chi$ and $r$ integrations. 

{\it Output.} In principle, the code can make available each of the scalar-field modes $\phi_{\ell m}$ anywhere in the computational domain. For our initial tests and for the purposes of his paper, we only output $\Phi^R$ and ${\cal F}^\alpha$ as functions of $t$ along the orbit, together with the final values $\delta\psi_{\rm cons}$ and $\delta\psi_{\rm diss}$.

The bulk of our code, including the numerical integrator of the field equation, is implemented in \texttt{C++}. However, presently there are several (computationally cheap) pre- and post-processing steps that are performed using \texttt{Mathematica}, for convenience. These include the calculation of the scattering trajectory, the interpolation of the self-force data along the orbit, and the numerical evaluation of the orbital integrals that yield $\delta\psi_{\rm cons}$ and $\delta\psi_{\rm diss}$. The latter numerical integration is performed using the default setting of \texttt{Mathematica}'s \texttt{NIntegrate} command.

\subsection{Error estimates}


The primary sources of numerical error in our calculation are from (1) finite-difference discretisation (finite $h$), (2) the truncation of mode sums at $\ell=\ell_{\rm max}$, (3) the truncation of the orbital integral at large $r$, (4) the interpolation of the discrete field and self-force data along the orbit, (5) the numerical integration along the orbit, and (6) the numerical integration of the geodesic equations. We have checked the influence of each of these errors on the final value of $\delta\psi$ by varying the appropriate numerical parameters. We have found that the dominant source of error varies in different regions of the parameter space.

In the strong field, the finite-difference discretisation introduces a relative error in $\delta\psi$ of the order $0.1\%$, with other errors at least an order of magnitude smaller. This level of accuracy means that error bars are too small to be visible on the scale of the sample strong-field orbit plots to be displayed in the next section.

As we increase the periastron distance $r_0$, we must also increase the final evolution radius $r_{\rm fin}$, and with it the initial radius $r_{\rm init}$ as well as the numerical evolution time (while resolution requirements remain roughly constant). At large periastron distance, our choice of $r_{\rm fin}$ (and $r_{\rm init}$) becomes restricted by our limited computational resources, and the error from the radial truncation becomes dominant. In the most extreme cases (largest $r_0$) displayed in the next section, this increases the relative error on $\delta\psi$ to a few percent.

A possible way to reduce finite-$r_{\rm fin}$ truncation error in weak-field orbits would be to analytically approximate the contribution to $\delta\psi$ from the truncated large-$r$ portion of the orbit, using the PM expressions for the self-force from Ref.\ \cite{Gralla2011}. We have tested this idea with some success, but have opted not to implement it here, in order to keep our PM comparisons truly independent. We do, however, use these analytic results to estimate the errors caused by the finite-$r_{\rm fin}$ truncation, which is  by far the dominant error in the weak-field regime. 


\section{Numerical implementation: sample results}
\label{Sec:results}

\begin{figure}[h!]
\centering
\includegraphics[width=\linewidth]{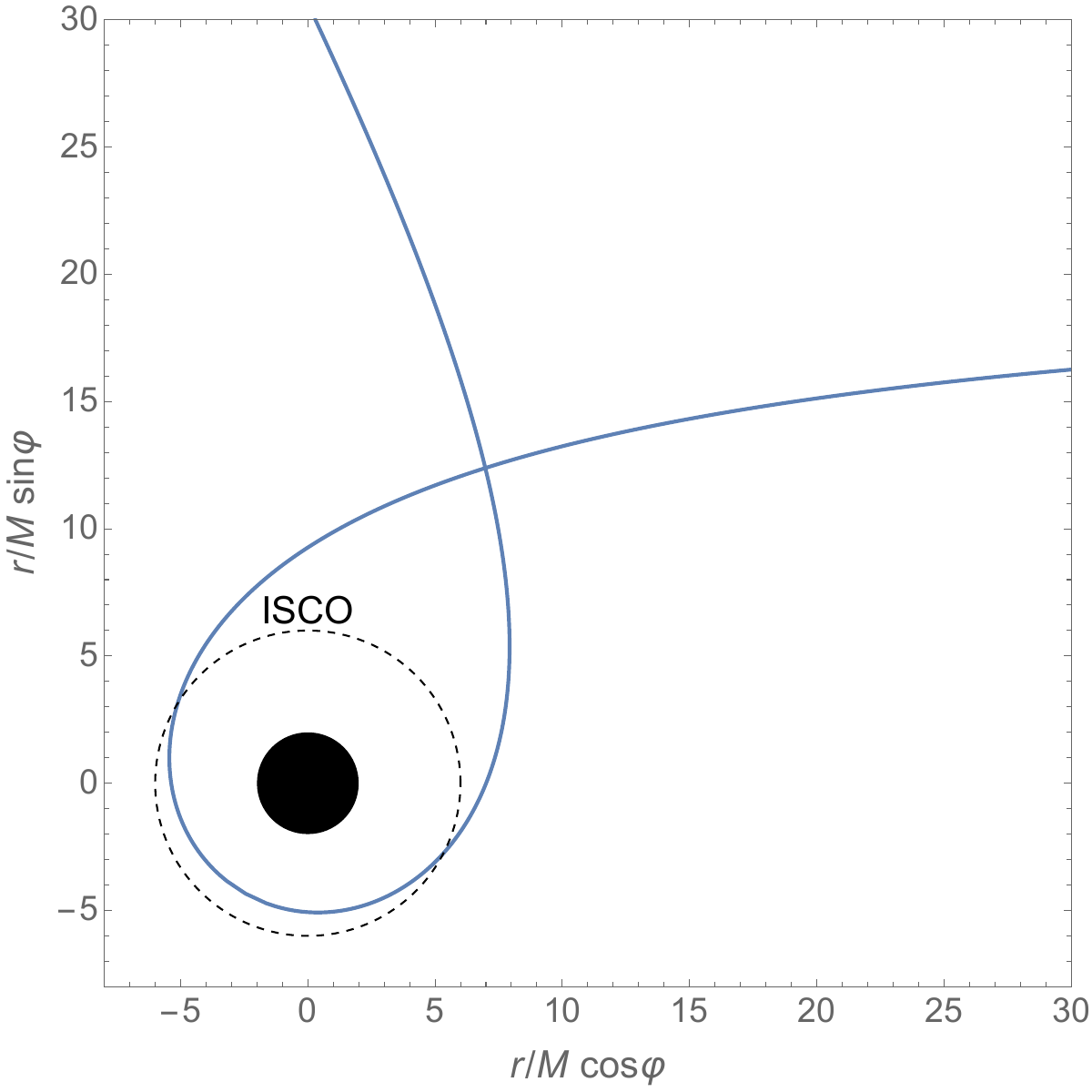}
\caption[A sample strong-field scattering geodesic orbit used for our numerical illustration]{A sample strong-field scattering geodesic orbit used for our numerical illustration, with parameters given in Eqs.\ (\ref{sampleorbit_vb}) and (\ref{sampleorbit}). The orbit is plotted in the equatorial plane using Cartesian-like coordinates $(x,y)=(r\cos\varphi,r\sin\varphi)$.  The location of the ISCO is shown for reference. Ignoring self-force, the scattering angle is $\psi \simeq 301^\circ$.}  
\label{orbit}
\end{figure}

Here we present a selection of numerical results for a scalar charge on a range of hyperbolic orbits. We consider first a typical ``strong-field'' scattering orbit, with geodesic parameters
\begin{equation}\label{sampleorbit_vb}
v = 0.2, \qquad b=21M,
\end{equation} 
corresponding to 
\begin{align}
r_0\simeq 4.98228M,\qquad  E\simeq&\: 1.02062,\qquad L\simeq 4.28661M, \nonumber \\
\quad e\simeq 1.1948, \qquad  p\simeq&\: 10.9351, \qquad \psi \simeq 301^\circ.
\label{sampleorbit}
\end{align}
The orbit is depicted in Fig.\ \ref{orbit}. Note that despite its seemingly large impact parameter, this geodesic orbit really is a strong-field one, with its periastron occurring well below the innermost stable circular orbit (ISCO), and with its large scattering angle $\psi$. 

\begin{figure}[htb]
\centering
\includegraphics[width=\linewidth]{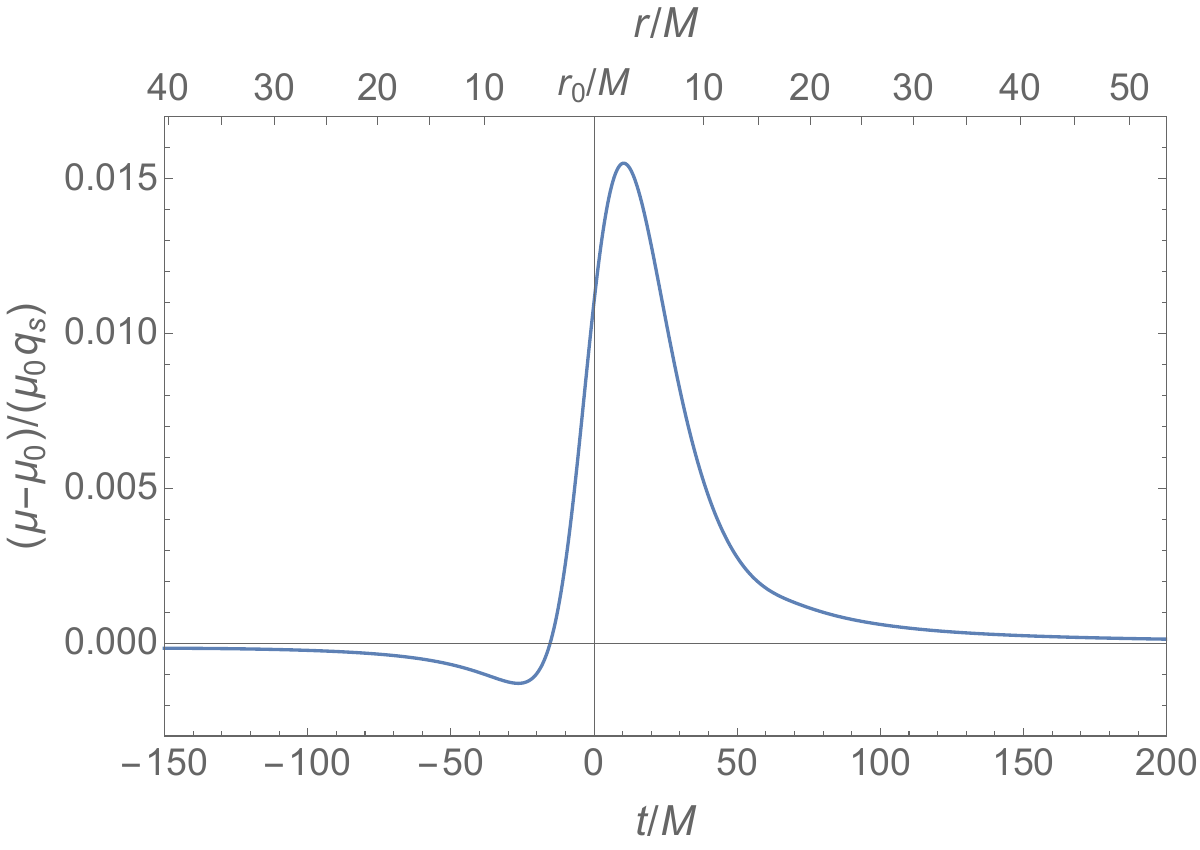}
\caption{The variation in the mass of the scalar particle (due to exchange of energy with the scalar field) along the orbit depicted in Fig.\ \ref{orbit}, as a function of time (lower scale) and orbital radius (upper scale). Shown is the relative mass difference [divided by the small dimensionless parameter $q_s=Q^2/(\mu_0 M)$] with respect to the mass $\mu_0$ at infinity. The periastron location at $t=0$ is indicated with a vertical line. 
}
\label{ScalarMassChange}
\end{figure}

Figure \ref{ScalarMassChange} shows the variation in the mass $\mu$ of the particle along the orbit depicted in Fig.\ \ref{orbit}, as calculated using Eq.\ (\ref{eqn:ScalarMass}). The maximal relative change in this case is $\sim 0.015q_s$, where, we recall, $q_s=Q^2/{\mu M}$ is the dimensionless small parameter of the scalar-charge model. Since the regular field $\Phi^R$ approaches zero at infinity, there is no net mass change overall. Notable features are the asymmetry in the magnitude of mass variation between the inbound and outbound legs of the orbit, and the small time lag between the peak mass and the periastron. Both features, we presume, can be attributed to retardation effect in the self-interaction. 

\begin{figure}[htb]
\centering
\includegraphics[width=\linewidth]{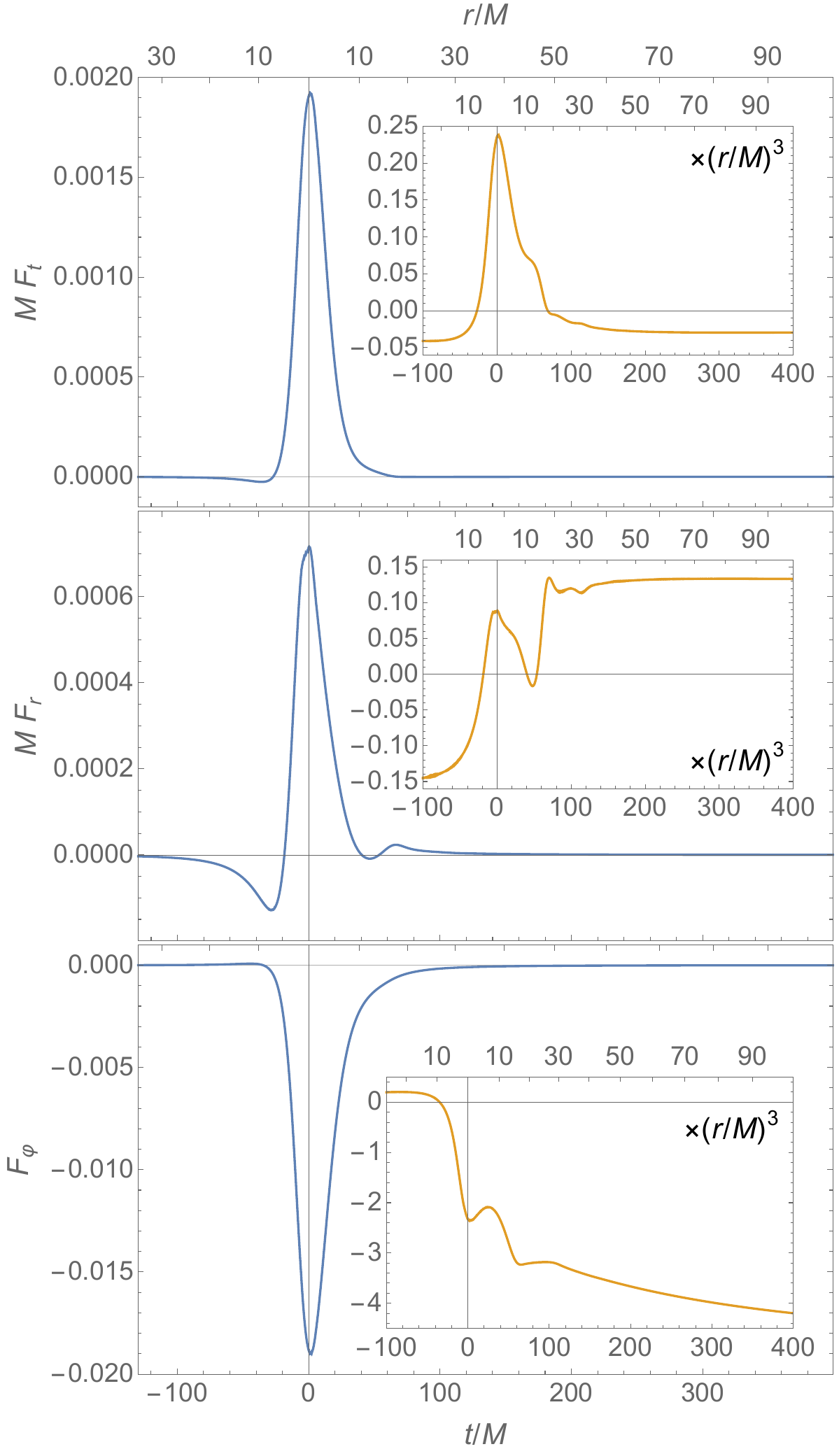}
\caption{The scalar-field self-force components $F_t$ ({\it top}), $F_r$ ({\it middle}) and $F_\varphi$ ({\it bottom}) along the orbit shown in Fig.\ \ref{orbit}, as functions of time $t$ (lower scale) and orbital radius $r$ (upper scale). The periastron location at $t=0$ is indicated with a vertical line. The insets show the same data rescaled by a factor $(r/M)^3$, which brings out more clearly the post-periastron oscillations associated with back-reaction from quasinormal ringing.}
\label{FPlots}
\end{figure}

In Fig.\ \ref{FPlots} we plot the self-force $F_\alpha$ along the geodesic orbit of Fig.\ \ref{orbit}. The self-force shows a similar lag (only slightly discernible in these plots) between the periastron and the peak amplitude. Another feature of note are the small undulations in the data a short time after periastron, visible more clearly in the insets. These are likely due to quasinormal-mode excitation, a phenomenon we have previously observed in the gravitational scattering problem \cite{LongBarack2021}. The same behavior had been studied in detail for highly eccentric bound orbits \cite{Nasipak:2019hxh,Thornburg:2019ukt}, where it was quantitatively associated with back-reaction from quasinormal ringing. The periastron lag has also been noted in previous work on bound orbits, e.g.~in \cite{Haas07}.

Given the self-force data, we can calculate the conservative and dissipative self-force corrections to the scattering angle using either 
Eqs.\ (\ref{deltaphi_methodI_cons}) and (\ref{deltavarphi_diss}) or (\ref{deltaphi1_fcons_method2}) and (\ref{deltaphi1_fdiss_method2}). We have done so for a large sample of geodesic orbits with parameters in the range $v\in [0.05,0.5]$ and $b\in [b_{\rm crit},150M]$. For each orbit we have applied both sets of formulas ($\chi$ integration and $r$ integration) for cross-validation. We have found that the results for $\delta\psi$ differed by $\sim 0.01\%$ at most, and typically much less; these differences are always smaller than other numerical errors in our calculation (to be discussed further below).  

\begin{figure}[htb]
\centering
\includegraphics[width=\linewidth]{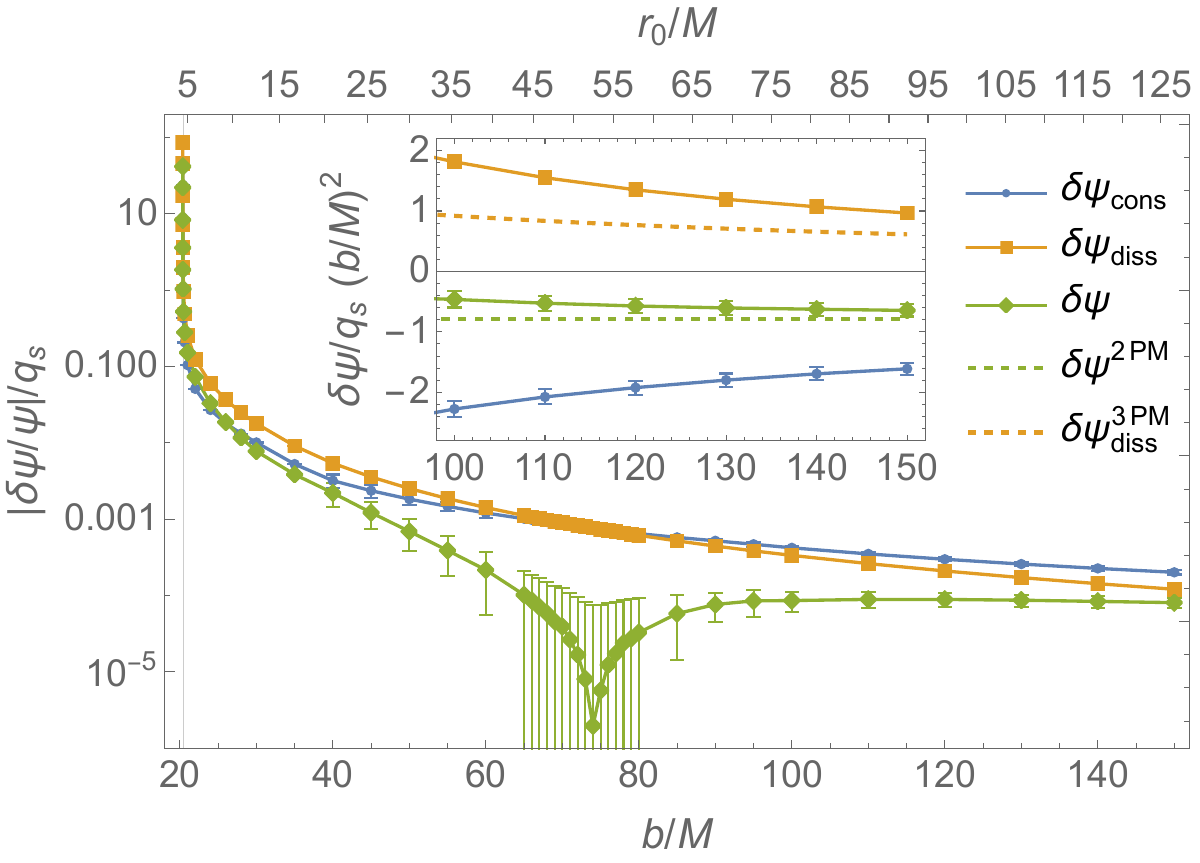}
\caption{The scalar-field self-force correction to the scattering angle for various orbits with $v=0.2$. We show here $\delta\psi/q_s$ as a fraction of the geodesic scattering angle $\psi$, as a function of impact parameter $b$ (lower scale) and periastron distance (upper scale). Displayed separately are the conservative contribution $\delta\psi_{\rm cons}$ (which is negative), the dissipative contribution $\delta\psi_{\rm diss}$ (which is positive), and the total $\delta\psi$ (which changes sign at around $b=74M$ for this value of $v$). The solid curves are interpolations through the numerical data points. The vertical line on the left represents the critical value of $b$, below which the orbit plunges into the black hole, $b_{\rm crit}\simeq 20.382M$. Note the self-force correction to the scattering angle blows up at $b_{\rm crit}$ faster than does the geodesic scattering angle; cf.\ Fig.\ \ref{bcrit} below. The inset shows a subset of large-$b$ data, as compared to the leading-order (2PM) conservative and (3PM) dissipative terms from Eqs.\ (\ref{PM_cons_scalar}) and (\ref{PM_diss_scalar}). Error bars (here and in all subsequent plots) are estimated from the magnitude of the truncated large-$r$ portion of the orbital integral, evaluated analytically using leading PM formulas (this is the dominant source of error wherever errors are visually discernible in our plots). Relative errors are large near the point where $\delta\psi$ changes it sign. 
}
\label{ScatAngleCorrv}
\end{figure}

\begin{figure}[htb]
\centering
\includegraphics[width=\linewidth]{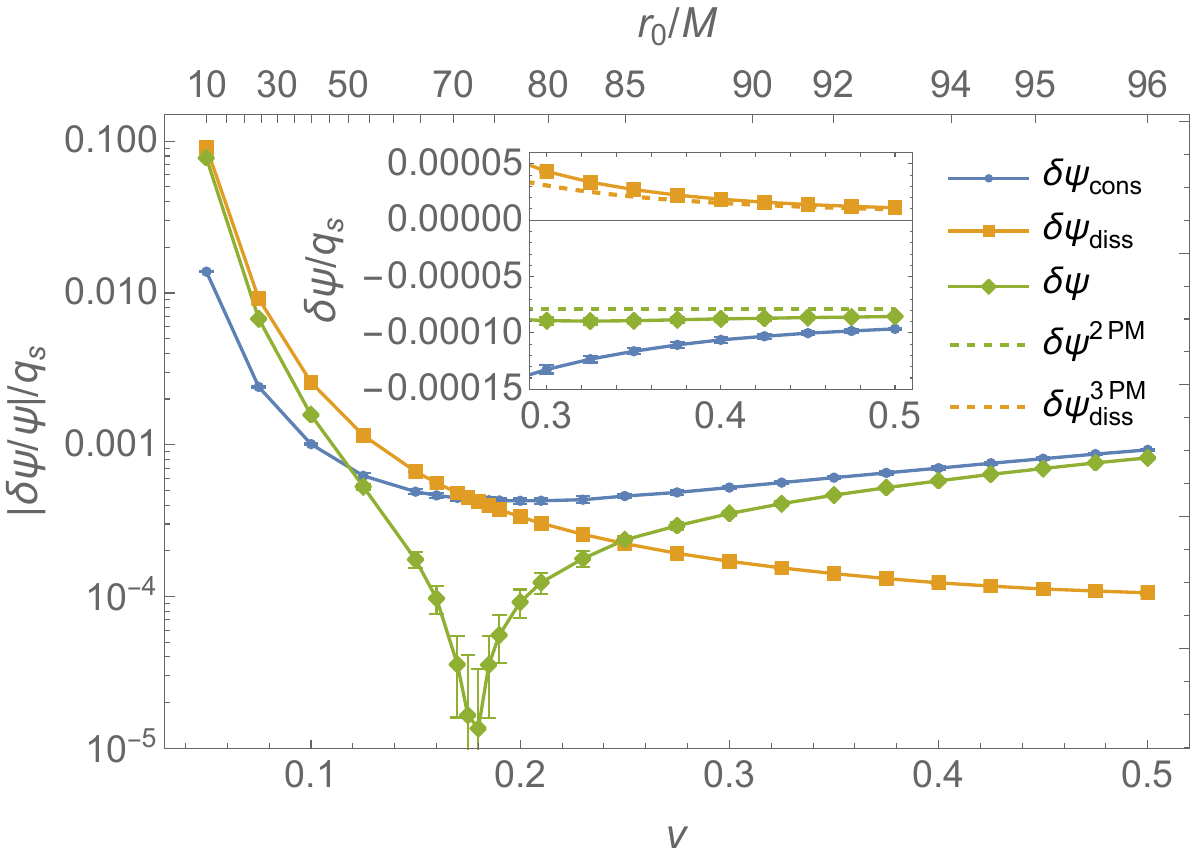}
\caption{
Similar to Fig.\ \ref{ScatAngleCorrv}, here showing results for a sample of orbits with fixed $b=100M$ and varying $v$.}
\label{ScatAngleCorrb}
\end{figure}

Figs.\ \ref{ScatAngleCorrv} and \ref{ScatAngleCorrb} show our numerical results for $\delta\psi_{\rm cons}$, $\delta\psi_{\rm diss}$ and the total $\delta\psi$ for a variety of orbits. In all cases we find $\delta\psi_{\rm cons}<0$ and $\delta\psi_{\rm diss}>0$: the conservative piece of the self-force {\em decreases} the scattering angle, while dissipation {\em increases} it. For weak-field orbits, the conservative effect [which is 2PM; recall Eq.\ (\ref{PM_cons_scalar})] dominates over the dissipative effect [which is 3PM; recall Eq.\ (\ref{PM_diss_scalar})], but the situation reverses for strong-field orbits, where the dissipative effect seems to dominate in general. As a result, the total correction $\delta\psi$ tends to be positive for close-approach encounters but negative in the weak-field regime. The function $\delta\psi(v,b)$ changes its sign in the transition between the two regimes. Figures \ref{ScatAngleCorrv} and \ref{ScatAngleCorrb} also illustrate how $\delta\psi_{\rm cons}$ and $\delta\psi_{\rm diss}$ seem to approach the corresponding leading-order PM values in expressions (\ref{PM_cons_scalar}) and (\ref{PM_diss_scalar}) in the weak-field limit, as expected.

\begin{figure}[htb]
\centering
\includegraphics[width=\linewidth]{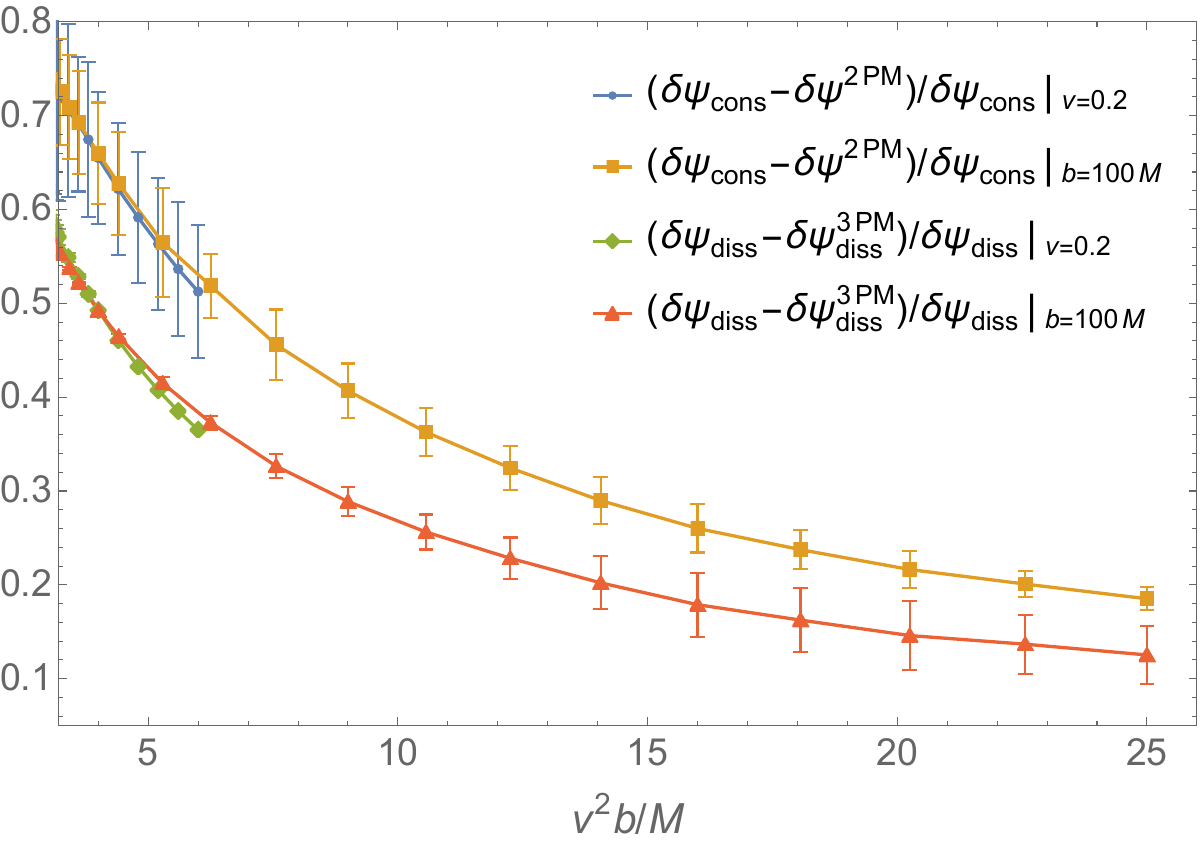}
\caption{Relative difference between (a subset of) the numerical $\delta\psi$ data shown in Figs.\ \ref{ScatAngleCorrv} and \ref{ScatAngleCorrb} and the leading-order analytical PM expressions from Eqs.\ (\ref{PM_cons_scalar}) and (\ref{PM_diss_scalar}). Solid curves are interpolations. 
}
\label{ScatAnglePMRelComp}
\end{figure}

Figure \ref{ScatAnglePMRelComp} shows the relative difference between the numerical data points and the leading-order PM values from Eqs.\ (\ref{PM_cons_scalar}) and (\ref{PM_diss_scalar}). For this plot we choose to adopt $v^2 b/M$ as a measure of how deep we are in the weak-field regime, recalling our discussion around Eq.\ (\ref{eqn:v^2b}). The agreement between the numerical data and the PM expressions becomes closer with increasing $v^2 b/M$, down to about $18\%$ and $12\%$ difference at $v^2 b/M\sim 25$ for the conservative and dissipative pieces respectively.  

\begin{figure}[htb]
\centering
\includegraphics[width=\linewidth]{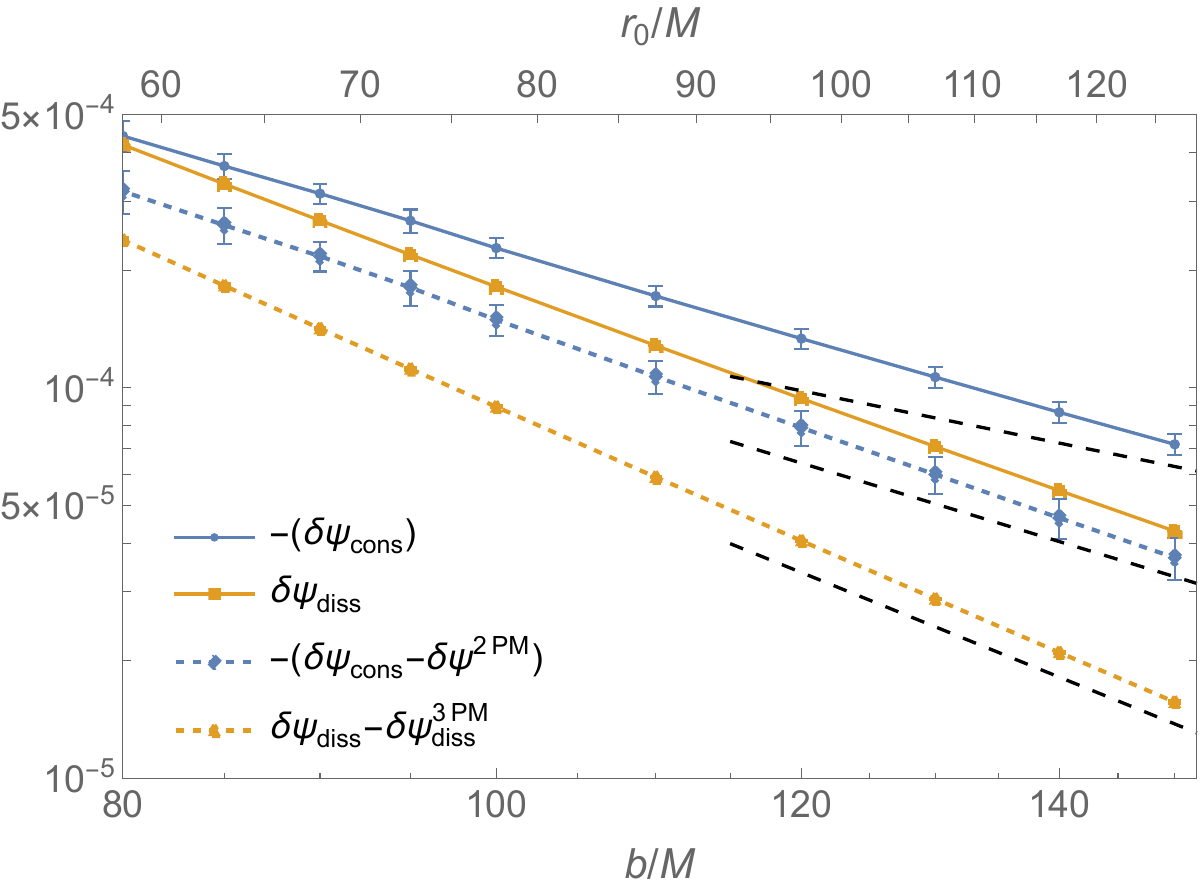}
\caption{A large-$b$ portion of the $v=0.2$ data displayed in Fig. \ref{ScatAngleCorrv}, here showing the (absolute) differences 
$-(\delta\psi_{\rm cons}-\delta\psi^{\rm 2PM})$ and $(\delta\psi_{\rm diss}-\delta\psi^{\rm 2PM})$ (per $q_s$, dashed), as well as $\delta\psi_{\rm cons}$ and $\delta\psi_{\rm diss}$ themselves (per $q_s$, solid), for reference. The long-dash straight lines are arbitrary reference curves $\propto b^{-2}$, $\propto b^{-3}$ and $\propto b^{-4}$ (top to bottom).}
\label{ScatAnglePMComp}
\end{figure}

To better quantify the weak-field behavior of our $\delta\psi$, Fig.\ \ref{ScatAnglePMComp} shows a large-$b$ segment of the $v=0.2$ data from Fig.\ \ref{ScatAngleCorrv}, along with the (absolute) difference between the numerical and leading-order PM values of $\delta\psi_{\rm cons}$ and $\delta\psi_{\rm diss}$. The plot roughly confirms the expected asymptotic scalings $(\delta\psi_{\rm cons}-\delta\psi^{\rm 2PM})\propto b^{-3}$ and $(\delta\psi_{\rm diss}-\delta\psi^{\rm 3PM}_{\rm diss})\propto b^{-4}$, but $\delta\psi_{\rm cons}$ itself appears to decay rather like $b^{-3}$ (instead of the expected $b^{-2}$). The likely explanation is that, for the values of $b$ shown, the behavior of the conservative sector is still dominated by subleading PM terms; the asymptotic $\propto b^{-2}$ behavior is expected to take over only at larger values of $b$, beyond our current reach for $v=0.2$. 

\begin{figure}[htb]
\centering
\includegraphics[width=\linewidth]{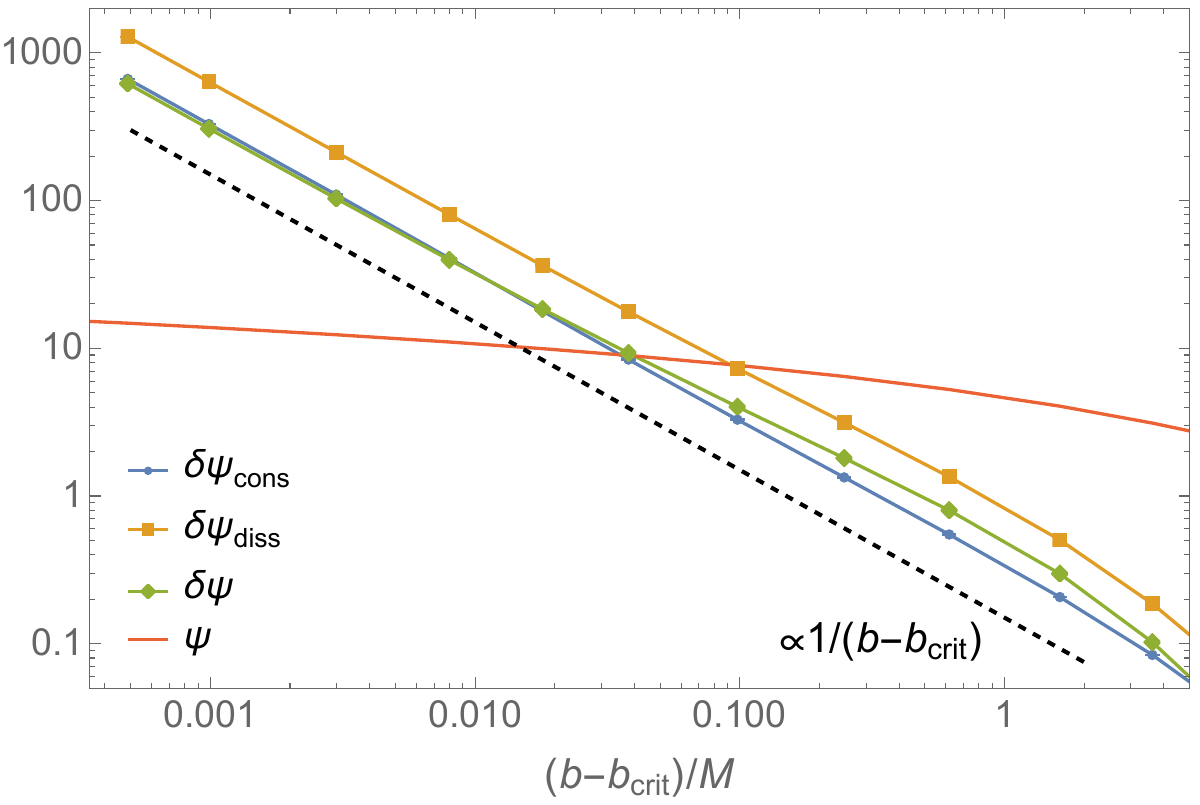}
\caption{The behavior of the self-force correction $\delta\psi$ at the approach to the critical orbit. While the geodesic scattering angle $\psi$ (shown for reference) diverges logarithmically with $b-b_{\rm crit}$, the self-force correction appears to diverge like $(b-b_{\rm crit})^{-1}$, and so do separately its conservative and dissipative pieces.}
\label{bcrit}
\end{figure}

A striking feature, manifest in Figs.\ \ref{ScatAngleCorrv} and \ref{ScatAngleCorrb}, is the rapid divergence of $\delta\psi$ (and of its separate conservative and dissipative pieces) at the approach to the critical orbit. Figure \ref{ScatAngleCorrv}, in particular, suggests this divergence is faster than that of the geodesic scattering angle $\psi$. Figure \ref{bcrit} explores this behavior in more detail. In the geodesic case, shown in the figure for reference, the divergence has the form 
\begin{equation}
\psi\propto \log(b-b_{\rm crit})
\end{equation}
 (at fixed $v$). This can be deduced analytically from the expressions in Sec.\ \ref{Sec:intro}; or see, for example, Eq.\ (106) of Ref.\ \cite{gund}.  The data in Fig.\ \ref{bcrit} suggest that the self-force correction, on the other hand, has the asymptotic behavior
\begin{equation}
\delta\psi \propto \frac{1}{b-b_{\rm crit}},
\end{equation}
and similarly for $\delta\psi_{\rm cons}$ and $\delta\psi_{\rm diss}$ in separate. With suitable additional numerical data it should be possible to fit for the $v$-dependent coefficient of this inverse-power divergence term. For a finite $q_s$, as we get closer to criticality, the self-force term becomes dominant and the self-force approximation breaks down.  

\section{Conclusion and outlook}
\label{Sec:conc}

In the formulation part of this work, Secs.\ \ref{Sec:SFMotion}--\ref{Sec:SFr}, we have developed general integral formulas for the self-force correction $\delta\psi$ to the scattering angle (at fixed $v,b$), given the self-force.  Equations (\ref{deltaphi_methodI_cons}) and (\ref{deltavarphi_diss}) give the conservative and dissipative pieces of this correction, $\delta\psi_{\rm cons}$ and $\delta\psi_{\rm diss}$, in terms of the $e,p$ parametrization of geodesic orbits and with the relativistic anomaly $\chi$ as an integration variable along the orbit. Equations (\ref{deltaphi1_fcons_method2}) and (\ref{deltaphi1_fdiss_method2}), alternatively, give $\delta\psi_{\rm cons}$ and $\delta\psi_{\rm diss}$ directly in terms of the parameters $v,b$, and with the radius $r$ as an integration variable. Both formulations may be useful in different circumstances, and in our implementation we have applied both to enable a cross-check. In Sec.\ \ref{sec:PMlimit} we have derived the leading-order PM reduction of our integral formulas, Eqs.\ (\ref{1PM_methodII_rad}) and (\ref{1PM_methodIIdiss_rad}), and confirmed that it returns the known PM results when applied with the leading-PM-order self-force available analytically from Ref.\ \cite{GrallaLobo2022}. 
(As a by-product of this weak-field test, we have analytically derived the leading, 3PM term of $\delta\psi_{\rm diss}$ in the scalar-charge model, which Ref.\ \cite{GrallaLobo2022} does not provide.)

Our formulation can be applied with the gravitational self-force to describe the physical problem of scattering of a mass particle off a Schwarzschild black hole. However, as it stands, it returns the scattering angle in an arbitrary frame inherited from the particular gauge in which the self-force is expressed. A natural way to remove this arbitrariness (and enable comparison with standard results in appropriate limits) would be to work out the (gauge) transformation of our expressions to the center-of-mass frame. Such an analysis could be modeled after the calculation done in Sec.\ V of Ref.\ \cite{Baracketal2019}, but we have not carried it out here, leaving it for future work. Further below we discuss other steps of development necessary to enable tackling the gravitational problem in full.  

In this work we proceeded to apply our formulation to the simpler physical model of a scattered scalar charge (neglecting the gravitational perturbation and the gravitational self-force), in which case no gauge ambiguity arises: the scattering angle is calculated in a Schwarzschild coordinate system centered at the center of the large black hole, which, however, coincides with the system's center of mass (since the gravitational mass of the particle is neglected). In this case our integral formulas can be applied as they are and without further correction, simply replacing the gravitational self-force with the (orthogonal component of the) back-reaction force from the scalar field. The scalar-field model was described in Sec.\ \ref{Sec:Scalar} and it was numerically implemented in Secs.\ \ref{Sec:ScalarImplementation} and \ref{Sec:results}. For our numerical implementation we took advantage of an existing numerical code, developed by us in \cite{LongBarack2021}, which required only minor adaptations. 
 
We have thus numerically calculated the scalar-field self-force and its correction to the scattering angle (and to the particle's mass) for a large sample of scattering orbits. As discussed in detail in Ref.\ \cite{LongBarack2021}, our code can comfortably handle strong-field scattering orbits in the approximate parameter range $v \lesssim 0.6$ and $r_0(v,b)\lesssim 150M$, but the computation becomes increasingly more demanding as we venture out to weaker-field orbits. (Note, however, that there is essentially no limitation on the value of $b$, provided that the corresponding periastron distance $r_0$ remains sufficiently small.) Using large-$r_0$ data we have been able to demonstrate a good agreement with leading-order PM results in both conservative and dissipative sectors. In forthcoming work we will report detailed comparisons with higher-order PM terms derived recently for the scalar-field model using quantum amplitude methods \cite{Bern_etal_inprep}. Our comparisons raises the prospect of being able to determine high-order PM terms of the scattering angle beyond those known analytically, by fitting to numerical self-force data, in much the same way this has previously been done in post-Newtonian theory. 

Of course, the main added value of the self-force approach is in its accurate description of strong-field features. With this in mind, we examined the behavior near the critical parameter-space surface separating scattering orbits from plunging ones. Our numerical results indicate that the self-force correction $\delta\psi$ exhibits a divergence $\propto(b-b_{\rm crit})^{-1}$, stronger than the logarithmic divergence of the geodesic $\psi$. A more detailed quantitative study of the near-separatrix behavior could in the future inform an efficient resummation of PM expressions to the effect of constructing a uniformly accurate analytical model of the scattering angle, in much the same way this was done (e.g.) in Ref.\ \cite{AkcayBarackDamour_etal2012} using the light-ring behavior to resum post-Newtonian expressions for Detweiler's redshift.  

The ultimate aim of our program is to perform similar calculations for the physical problem of pure-gravity scattering. As mentioned, this will require a careful consideration of the gauge ambiguity inherent in the gravitational problem. In particular, a suitable transformation to the center of mass would need to be devised and applied to our expressions. 

In parallel, an appropriate numerical technology would need to be developed for calculating the metric perturbation from scattering orbits in a gauge appropriate for self-force calculations.  A main step towards this goal was taken by us in Ref.\ \cite{LongBarack2021}, where we have formulated a metric reconstruction procedure for scattering orbits and illustrated its numerical implementation. Our method is based on a numerical time-domain evolution of the Teukolsky equation for a certain scalar-like Hertz potential, from which the metric perturbation is obtained by applying a second-order differential operator. The particular, basic numerical evolution method applied in Ref.\ \cite{LongBarack2021} (similar to the method used in the current work) turned out to be susceptible to instabilities associated with certain nonphysical growing-mode solutions of the Teukolsky equation. These required us to implement certain remedies that incurred heavy computational overhead, unfortunately. To overcome this problem, we suggested in Ref.\ \cite{LongBarack2021} the use of a numerical evolution method based on hyperboloidal slicing with compactification (of the like of the methods developed, e.g., in \cite{Racz:2011qu,Zenginoglu:2012us,Harms2013,Macedo2014,CsukasRacz2019,CsukasRacz2021,Macedo2022}), which, we argued, should be inherently immune to the problem of growing modes. We are currently working to develop a suitable code \cite{MacedoLongBarackinprog} based on the ideas introduced in Ref.\ \cite{Macedo2014}.  

The proposed numerical method, like our method in this work, is based on an integration of the relevant field equations in the {\em time domain}. This is a natural strategy in the scattering problem, where (unlike in the case of bound orbits) the field admits a continuous spectrum. However, the approach involves solving partial differential equations, which is computationally intensive and (consequently) produces results of limited numerical precision. An alternative approach would be based on a full Fourier-harmonic decomposition of the relevant field equations (e.g., the Teukolsky or the Klein-Gordon equations), which would reduce the numerical task to the solution of {\em ordinary} differential equations. Such a frequency-domain approach is the mainstay of self-force calculations for bound orbits \cite{vandeMeent2018}, but it is yet to be fully developed for scattering orbits, where the continuous spectrum and slowly converging Fourier integrals pose new challenges. Preliminary results suggest that a frequency-domain approach has the potential to dramatically increase the precision of self-force calculations for scattering orbits \cite{WhittallBarack_inprep}. Such improved precision would be crucial, for instance, in a program to extract high-order PM parameters.

\section*{Acknowledgments}

We are grateful to Maarten van de Meent for introducing us to the simple-form solutions in Eqs.\ (\ref{r0})--(\ref{r2}), and to Zvi Bern for helpful comments on a draft of this work. 
OL acknowledges support from EPSRC through Grant Nos.\ EP/R513325/1 and EP/T517859/1. This research was supported in part by the National Science Foundation under Grant No. NSF PHY-1748958. We acknowledge the use of the IRIDIS High Performance Computing Facility, and associated support services at the University of Southampton, in the completion of this work. This work makes use of the Black Hole Perturbation Toolkit \cite{BHPToolkit}.

\appendix

\section{Convergence of the integral in Eq.\ (\ref{deltavarphi_sf})}\label{App:Convergence}

The purpose of this appendix is to establish the convergence of the integral over $\chi$ in the expression (\ref{deltavarphi_sf}) for the self-force correction $\delta\psi$ to the scattering angle.  

Consider first the behavior near the limits $\chi\to \pm \chiinf$. From Eqs.\ (\ref{tauchi}) and (\ref{fEfL}) we have $\tau_{\chi}\sim (\chi\mp\chiinf)^{-2}$ and $f_E,f_L\sim (\chi\mp\chiinf)^{0}$. Therefore, assuming the self-force components $F_t$ and $F_\varphi$ fall off at infinity (which our numerical results confirm), each of the two integrals over $\chi'$ in Eq.\ (\ref{deltavarphi_sf}) either converges at $\chi\to\pm\chiinf$ or it diverges there slower than $\sim (\chi\mp\chiinf)^{-1}$. It follows that the final integral over $\chi$ converges at $\chi\to \pm \chiinf$.

Consider next the behavior near the periastron, $\chi=0$, which is more subtle. Here $\tau_{\chi}$ is bounded and nonzero, $F_t$ and $F_\varphi$ also bounded and nonzero, but $f_E,f_L \sim \chi^{-2}$. Hence the integral over $\chi$ of the separate $F_t$ and $F_\varphi$ terms actually diverges (logarithmically) at $\chi=0$. We can verify, however, that the integral over the sum of two terms is in fact convergent: the singular term cancels out between these two terms. 

To see this, we use the $|\chi|\ll 1$ (near-periastron) expansions
\begin{eqnarray}
f_E &=& -\frac{p\sqrt{p-3-e^2}\sqrt{(p-2)^2-4e^2}}{e^2(p-6-2e)^{3/2}}\, \chi^{-2} + O(\chi^0),
\nonumber\\
f_L &=& \frac{\sqrt{p-3-e^2}(p-2-2e)(1+e)^2}{Me^2\sqrt{p} (p-6-2e)^{3/2}} \chi^{-2} + O(\chi^0),
\nonumber\\
\tau_\chi &=& \frac{M p^{3/2}\sqrt{p-3-e^2}}{(1+e)^2 \sqrt{p-6-2e}} +O(\chi^2) , 
\nonumber\\
u^t & = & \sqrt{\frac{p(p-2+2e)}{(p-2-2e)(p-3-e^2)}} +O(\chi^2) ,
\nonumber\\
u^\varphi & = & \frac{(1+e)^2}{M p \sqrt{p-3-e^2}} +O(\chi^2) ,
\end{eqnarray}
to obtain 
\begin{widetext}
\begin{align}\label{convergence}
f_E(\chi)\int_{0}^{\chi} F_t(\chi')\tau_{\chi'}d\chi' 
&- f_L(\chi)\int_{0}^{\chi} F_\varphi(\chi') \tau_{\chi'}d\chi'
\nonumber\\
&= -\frac{M p (p-3-e^2)(p-2-2e)}{e^2(1+e)^2(p-2e-6)^2}\left(\sqrt{\frac{p(p-2+2e)}{p-2-2e}}\, F_t(0)+\frac{(1+e)^2}{M p} \, F_{\varphi}(0)\right) \chi^{-1} +O(\chi^0)
\nonumber\\
&= -\frac{M p (p-3-e^2)^{3/2}(p-2-2e)}{e^2(1+e)^2(p-2e-6)^2}\left[u^t(0) F_t(0)+u^\varphi(0) F_{\varphi}(0)\right] \chi^{-1} +O(\chi^0) .
\end{align}
\end{widetext}
The expression in square brackets in the last line is simply $-u^r(0) F_r(0)$, by virtue of the orthogonality relation $u^\alpha F_\alpha =0$. But $u^r(0)=0$, so the $O(\chi^{-1})$ term in Eq.\ (\ref{convergence}) drops, and we find that the entire expression is bounded. Thus the full integrand of the $\chi$ integral in Eq.\ (\ref{deltavarphi_sf}) is bounded at $\chi=0$, and the integral over $\chi$ converges there.

\section{Finite-difference scheme}
\label{app:FDS}

In this appendix we detail the finite-difference (FD) scheme used to solve the 1+1D sourced scalar-field equation (\ref{eqn:SourcedFieldEquation}). The equation has the form
\begin{equation}
\phi_{,uv} +\frac{1}{4}U(r) \phi = S_\phi,
\label{eqn:FDSFieldEquation}
\end{equation}
where $u,v$ are Eddington--Finkelstein null coordinates, the potential $U(r)$ is given in (\ref{eq:ScalarPotential}), and the distributional source can be read off (\ref{eqn:SourcedFieldEquation}). Our derivation follows the method of Ref.\ \cite{BarackSago2010} (which itself expands on a long history of work in constructing time-domain FD schemes for self-force applications, e.g.\ \cite{Lousto05, Haas07}).

\begin{figure}[htb]
\centering
\includegraphics[width=0.7\linewidth]{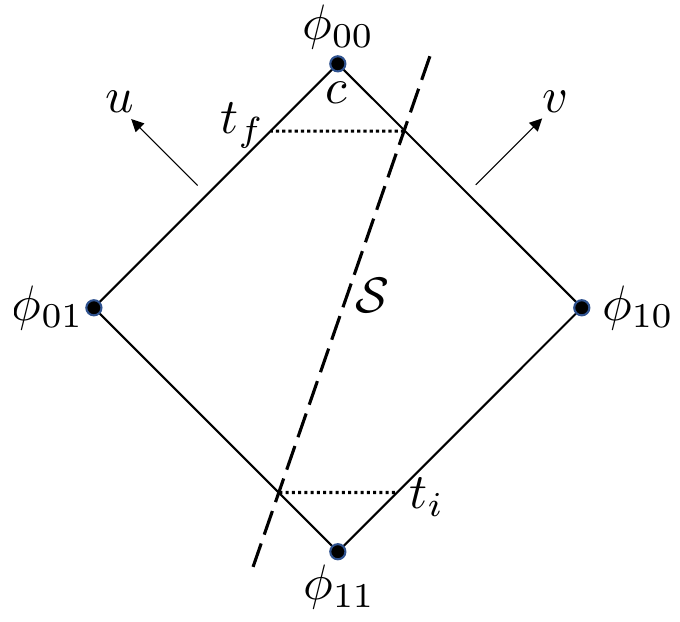}
\caption{A particle cell is traversed by the particle's worldline $\cal S$ (dashed curve). The apex of the cell is the point $c$ at $(u,v) = (u_c,v_c)$, and, in reference to it, we denote by $\phi_{nk}$ the numerical field values at a grid point with coordinates $(u,v)=(u_c- nh,v_c-kh)$. The particle enters (exits) the cell at time $t=t_i$ ($t=t_f$), which are calculated and stored in advance of the numerical evolution.}
\label{ScalarGenericCell}
\end{figure}

Recall our 1+1D numerical grid, shown in Fig.\ \ref{uvGrid}, which is constructed of uniform cells of size $h\times h$ in $u,v$ coordinates. Consider an arbitrary grid point $c$ with coordinates $(u,v)=(u_c,v_c)$. We denote by $\phi_{nk}$ the value of the numerical field $\phi$ at coordinates $(u,v)=(u_c- nh,v_c-kh)$, as illustrated in Fig.\ \ref{ScalarGenericCell} for a grid cell intersected by the particle's worldline $\cal S$. Our aim is to obtain a FD formula for the field at $c$, $\phi_{00}$, given the values $\phi_{nk}$ for all $n,k>0$, which are assumed known from previous steps in the characteristic evolution. Our goal is a scheme with a global quadratic convergence, i.e.\ an accumulated error in $\phi$ that scales as $h^2$. Since the total number of grid points scales as $h^{-2}$, we require, in general, a local (single-point) FD error no larger than $O(h^4)$.

In reference to a grid cell $C$ (with top vertex $c$), we distinguish between two cases: 
(1) the particle's worldline is external to the integration cell (``vacuum cell''), or (2) $\cal S$ passes directly through $C$ (``particle cell''), as shown in Fig.\ \ref{ScalarGenericCell}. We consider these two scenarios separately below. 

\subsection{Vacuum cells}
\label{app:FDSVac}

First we consider the scenario where $\cal S$ does not cross the integration cell. It is sufficiently accurate to write the FD approximation for $\phi_{00}$ based only on the three values $\phi_{01}$, $\phi_{10}$ and $\phi_{11}$. Integrating the two terms on the left-hand side of Eq.\ (\ref{eqn:FDSFieldEquation}) over the grid cell $C$ gives
\begin{equation}
\int_C \phi_{,uv} \: du dv = \phi_{00} - \phi_{01} - \phi_{10} + \phi_{11}
\end{equation}
(exactly), and 
\begin{equation}
\int_C \frac{1}{4}U(r)\phi \: du dv = \frac{1}{8} h^2U(r_c) \left( \phi_{01} + \phi_{10} \right) + O(h^4),
\end{equation}
where 
$r_c$ is the value of $r$ at point $c$. The homogeneous version of Eq.\ (\ref{eqn:FDSFieldEquation}) then yields
\begin{equation}
\phi_{00} = - \phi_{11} + (\phi_{01} + \phi_{10}) \left( 1 - \frac{h^2}{8} U(r_c) \right)+O(h^4),
\label{eqn:FDSVac}
\end{equation}
which is our FD formula for vacuum cells.

\subsection{Particle cells}
\label{app:FDSSource}

The vacuum formula (\ref{eqn:FDSVac}) does not work for cells that are traversed by the worldline, since there is then also a contribution from the distributional source $S_\phi$. Integrating the sourced equation (\ref{eqn:FDSFieldEquation}) over the cell, we obtain
\begin{equation}
\phi_{00} = \: - \phi_{11} + (\phi_{01} + \phi_{10}) \left( 1 - \frac{h^2}{8} U(r_c) \right)+ Z +O(h^3).
\label{eqn:FDSSourceGeneric}
\end{equation}
Here we have
\begin{eqnarray}\label{eqn:FDSSourceInt}
Z&=&\int_C S_\phi \: dudv
\nonumber\\
&=& \int_{t_i}^{t_f} \frac{f(r_p(t))}{E r_p(t)} \bar{Y}_{\ell m}(\pi/2, \varphi_p(t)) \: dt,
\end{eqnarray}
where we have recalled the explicit form of the source from Eq.\ (\ref{eqn:SourcedFieldEquation}),
and where $t=t_i$  and $t=t_f$ are the times at which the particle enters and exits the cell, respectively, as illustrated in Fig.\ \ref{ScalarGenericCell}. We cannot evaluate this integral analytically in exact form, but we can do so approximately at the required order in $h$. To this end, we choose to expand the integrand of Eq.\ (\ref{eqn:FDSSourceInt}) in $t$ about the time  $t_C=(t_i+t_f)/2$, midway between $t_i$ and $t_f$. Expanding thus to $O(t-t_C)$ and evaluating the integral, we obtain 
\begin{equation}\label{Z}
Z= \frac{f(r_C)}{E r_C}\bar{Y}_{\ell m}(\pi/2, \varphi_C)(t_f-t_i) + O(h^3),
\end{equation}
where $r_C:=r_p(t_C)$ and $\varphi_C:=\varphi_p(t_C)$.
The $O(h^3)$ cell error here is larger than the $O(h^4)$ for a vacuum cell, but it is permissible for us, since the number of particle cells scales only as $\sim h^{-1}$: an $O(h^3)$ local error in particle cells accumulates to give an $O(h^2)$ global error, still consistent with our requirement for a quadratic convergence. 

In summary, our second-order-convergent FD scheme is described in Eq.\ (\ref{eqn:FDSVac}) for vacuum cells, and in  Eq.\ (\ref{eqn:FDSSourceGeneric}) with Eq.\ (\ref{Z}) for particle cells.

\bibliographystyle{unsrt}
\bibliography{biblio}

\end{document}